\documentclass[12pt]{article}
\usepackage[utf8]{inputenc}
\usepackage{amsmath}
\usepackage{amssymb}
\usepackage{wasysym}
\usepackage{appendix}
\usepackage{graphicx}
\usepackage{hyperref}
\usepackage{listings}
\usepackage{color}
\usepackage{tcai}
\usepackage{array}
\usepackage{geometry}
\usepackage{amsthm}
\usepackage{natbib}
\usepackage{authblk}
\usepackage{setspace}
\usepackage{booktabs}
\usepackage{multirow}
\usepackage{mathtools}
\usepackage{bigstrut}

\newgeometry{left=1in,right=1in,bottom=1.5cm,top=1.0cm}

\title{Bayesian Mixed Effects  Models  for  Zero-inflated Compositions in Microbiome Data Analysis}
\author[1,$\dagger$]{Boyu Ren}
\author[2]{Sergio Bacallado}
\author[3]{Stefano Favaro}
\author[4]{Tommi Vatanen}
\author[1,4,$\ast$]{Curtis Huttenhower}
\author[1,$\ast$]{Lorenzo Trippa}
\affil[1]{Harvard T.H. Chan School of Public Health, Boston, USA}
\affil[2]{University of Cambridge, Cambridge, UK}
\affil[3]{Universit\`a degli Studi di Torino and Collegio Carlo Alberto, Turin, Italy}
\affil[4]{Broad Institute, Cambridge, USA}
\affil[$\dagger$]{\small{\textit{email:} bor158@mail.harvard.edu}}

\begin{document}

\maketitle

\begin{abstract}
Detecting associations between microbial compositions and sample characteristics is one of the most important tasks in microbiome studies. Most of the existing methods apply univariate models to single microbial species separately, with adjustments for multiple hypothesis testing. We propose a Bayesian analysis for a generalized mixed effects linear model tailored to this application. The marginal prior on each microbial composition is a  Dirichlet Process, and dependence across compositions is induced through a linear combination of individual covariates, such as disease biomarkers or the subject's age, and latent factors. The latent factors capture residual variability and their dimensionality is learned from the data in a fully Bayesian procedure. The  proposed model  is  tested  in  data analyses  and    simulation  studies  with zero-inflated compositions.  In  these  settings, within      each  sample,  a large proportion  of  counts  per  microbial  species  are equal  to  zero.  In our Bayesian model  \textit{a  priori}  the  probability  of    compositions  with  absent   microbial  species  is  strictly  positive. We propose an efficient algorithm to sample from the posterior and visualizations of model parameters which reveal associations between  covariates and microbial compositions.  We  evaluate the proposed method in simulation studies, and  then analyze a microbiome dataset  for infants with type 1 diabetes which  contains a large  proportion of zeros in the sample-specific microbial compositions.
\end{abstract}

\noindent%
\vfill

\newpage

\section{Introduction}
Large scale studies of the human microbiome have become increasingly common thanks to advances in next generation sequencing (NGS) technologies \citep{MetaHIT, HMP}. 
A  relevant task in these studies is to measure the association between a sample's microbial composition and  individual characteristics, such as biomarkers  and aspects of the sample's environment \citep{xochi2012,quince2013impact,kostic2015dynamics}. 
The abundances of microbial taxa are measured by assigning DNA reads  to reference genomes. Some experiments target specific genes, such as the 16S rRNA gene, while others sample the entire bacterial genome. In all cases, the resulting count data for a collection of samples are organized into a contingency table known as the operational taxonomic unit (OTU) table. 

Several methods for association studies with microbial data apply ideas from RNA-seq and other high-throughput genomic experiments \citep{edger,deseq, metagenomeseq}. 
These methods use raw or transformed counts of microbial species to test the association of a single species with relevant covariates. Typically, these tests are carried out one species at a time by using generalized linear models (GLMs) combined with families of distributions that are over-dispersed and zero-inflated \citep{xu2015} to accommodate well-known characteristics of microbial abundance data \citep{HZLi-review}. The major drawback of this approach is that it models species independently. This approach does not take into account correlations across microbial species and does not allow borrowing of information across species.

The outlined limitation has prompted the introduction of joint models of microbial abundance \citep{multinomial-dirichlet, logistic-normal, wadsworth2017integrative, MIMIX}. These methods model the counts of $I$ microbial species $(n_{i,j};i=1,\ldots,I)$, of a specific sample $j$, say a saliva sample, with a multinomial distribution. 
To account for the overdispersion these methods  assume the multinomial parameter $P^j=(P_{1}^j,\ldots,P_{I}^j)$ is random and distributed accordingly to a parametric model. For example, in \cite{multinomial-dirichlet} and \cite{wadsworth2017integrative}, $P^j$'s follow independent Dirichlet distributions and in \cite{logistic-normal} and \cite{MIMIX}, $P^j$'s follow  multivariate logistic-normal distributions.
To associate the covariates to the microbial compositions, all models link the parameters of each distribution of $P^j$  (Dirichlet or  logistic-normal) to  covariates of sample $j$ via a regression function. Inference on the regression coefficients indicates whether a covariate is associated with the abundance of a species or not. 
Although these joint models overcome  limitations of separate modeling of single species, the assumed distributions of the $P^j$'s in these methods are restrictive. For instance, $P_i^j$ is strictly positive for all $i$ and $j$. This does not reflect the fact that some species can be completely absent in sample $j$. In addition, the variation of $P^j$'s across samples might be mainly associated to some latent characteristics that are not observed. In this case, the methods which link  model parameters exclusively to covariates do not capture dependence  across  species-specific  residuals.

Bayesian nonparametric methods that jointly model the compositions $P^j$ offer flexible alternatives. A widely used class of nonparametric models stems from the Hierarchical Dirichlet process \citep{hdp}. In its simplest form, the Hierarchical Dirichlet process (HDP) assumes samples are exchangeable and the compositions  $P^j$  over  $I=\infty$ species are  identically  distributed. The exchangeability assumption in the HDP does not capture potential association between $P^j$'s and covariates. Nonparametric models with covariates explicitly embedded are ideal candidates for modeling dependence of compositions $P^j$ on covariates. There are only a few such models discussed in literature. A relevant class of nonparametric models embedding covariates utilizes the Chinese restaurant processes representation \citep{johnson2013bayesian}. A second class of such models utilizes completely random measures \citep{lijoi2014bayesian}. A third class of models follows the idea in \cite{DDP}. Among this class of models, \cite{rodriguez2011nonparametric}, \cite{muller2011product}, \cite{griffin2013comparing}, and \cite{arbel}  construct dependent random measures using stick-breaking processes with atoms and weights  specified  through covariate-indexed stochastic processes. 

Recently, a Bayesian nonparametric model for microbiome data specified through sample-specific latent factors has been discussed in \cite{boyuren}. This construction induces a marginal Dirichlet process prior for each composition $P^j$ and introduces dependences across samples by associating microbial compositions $P^j$ to linear combinations of  latent factors. In addition, the authors introduced a link function with hard-truncation at zero to model zero-inflation in microbiome data. This model employs a shrinkage prior on the latent factors to produce parsimonious estimates that concentrate on a low-dimensional space.

This manuscript builds on the model of \cite{boyuren}, linking the microbial composition $P^j$ to covariate effects as well as to the latent factors. The resulting extended model takes into account  overdispersion and zero-inflation in microbiome data. 
More importantly, it can also enable association studies for microbiome data with efficient computations. 
By estimating coefficients for  linear combinations of relevant covariates, we visualize and infer whether a given covariate is associated with the microbial compositions or not. 

We performed an extensive simulation analysis to compare the performances  of the proposed  model and a parametric model with latent factors \citep{MIMIX} that is  used in microbiome studies. The simulation study suggests that our model can accurately recover population-level trends of microbial abundances over covariates even when the model is misspecified. Our model has better performance than \cite{MIMIX} in estimating the relationship between covariates and microbial abundances when the level of zero-inflation in the data increases. We also discuss the interpretation of model parameters and propose approaches to visualize covariates' effects.

The paper is organized as follows. In Section \ref{sec:2}   we  specify   the  Bayesian model and  discuss the identifiability of  relevant  model parameters. Section \ref{sec:3}  is  dedicated  to  computational aspects  and   provides  an overview of the sampling algorithm for  posterior  inference. Section \ref{sec:4}  presents simulation studies and in Section \ref{application}  we  discuss an application of the model to  data  from  type 1 diabetes studies  which collected longitudinal   measurements from a  cohort of infants. 
Section \ref{sec:6} concludes and discusses possible  extensions  of  the proposed analyses.

\section{Prior  model}
\label{sec:2}
In this section, we  first  review the construction of the Dependent Dirichlet processes in \cite{boyuren}, and then provide a new version of the model which incorporates covariates. We also discuss the identifiability of the model parameters,  including  the parameters that correspond to the  covariates' effects. 
The model  will  be used in the next sections to analyze the OTU table $\bn=(n_{i,j}; i\leq I,j\leq J)$, where $n_{i,j}$ is the observed count of the microbial species $i$ in sample $j$. $I$ and $J$ are the total number of species and samples respectively. Our aim is  to extract  from  the OTU table information on  the relationships between microbial composition and observed samples' characteristics.

\subsection{Dependent Dirichlet processes}
\label{prev.model}

In Table \ref{otu.table.example}, we illustrate  a subset of  the OTU table from the DIABIMMUNE project \citep{tommi}. The goal of the DIABIMMUNE project is to compare microbiome communities in
infants with type 1 diabetes (T1D) or serum auto-antibodies (markers predicting
the onset of T1D) and healthy controls in three countries: Finland (FIN), Estonia (EST) and Russia (RUS). The study is
prospective and longitudinal, and the microbial abundances are measured with shotgun sequencing. Table \ref{otu.table.example} records  the  counts of 10 microbial species in three Russian samples and three Finnish samples based on 16S rRNA sequencing. We denote the $i$th recorded species by $Z_i$. For instance, $Z_1$ is Bifidobacterium longum in Table \ref{otu.table.example}.

For sample $j$, we assume the vector $(n_{1,j},\ldots,n_{I,j})$ follows a multinomial distribution with unknown parameters. 
Our analyses extend easily to the case in which the counts $n_{i,j}$  are Poisson random variables with unknown means. 
The sequencing  depth $n_j=\sum_{i=1}^I n_{i,j}$ and the sample-specific multinomial probabilities $(P_{1}^j,\ldots,P_{I}^j)$  determine the  distribution of $(n_{i,j};i\leq I)$. The probabilities $(P_{1}^j,\ldots, P_{I}^j)$ represent the microbial composition of sample $j$. We use $P^j(\{Z_i\})=P_{i}^{j}$ to denote the relative abundance of $Z_i$ in sample $j$. The vectors $P^j$ vary across samples according to  heterogeneity of  either  measured or  unknown  characteristics of the $J$ samples. For example, in Table \ref{otu.table.example},  the maximum  likelihood  estimates  (MLE)  of abundances of Bifidobacterium longum ($P^j(\{Z_1\})$) tend to  be  higher in Russian samples than in the Finnish samples. 

\begin{table}
\footnotesize
\centering
\caption[A subset of DIABIMMUNE dataset]{{An example of OTU table \citep{tommi}.}}
\label{otu.table.example}
\begin{tabular}{@{}ccccccccccccc@{}}
\toprule
Species                      & RUS1  & RUS2 & RUS3  & FIN1  & FIN2   & FIN3  \\
\midrule
Bifidobacterium longum       & 0       & 73222  & 3014074 & 14294   & 7291     & 9228    \\
Bifidobacterium bifidum      & 3594189 & 49223  & 0       & 11177   & 11656816 & 14759   \\
Escherichia coli             & 4210380 & 23025  & 635855  & 29700   & 7508     & 556208  \\
Bifidobacterium breve        & 0       & 136    & 245827  & 19312   & 7223273  & 0       \\
Bacteroides fragilis         & 0       & 88751  & 0       & 6257732 & 343      & 75506   \\
Bacteroides vulgatus         & 0       & 7454   & 0       & 4745    & 0        & 25859   \\
Bacteroides dorei            & 0       & 0      & 0       & 0       & 0        & 0       \\
Bifidobacterium adolescentis & 0       & 111248 & 1626357 & 735715  & 1194     & 0       \\
Bacteroides uniformis        & 0       & 3901   & 0       & 5859    & 1633     & 28638   \\
Ruminococcus gnavus          & 145485  & 33004  & 92101   & 253830  & 29       & 1186774 \\
\bottomrule
\end{tabular}
\end{table}

We describe the  Bayesian model for  the unknown compositions $P^j$, $j=1,\ldots,J$ in \cite{boyuren}. 
Let  $\Zsc$ be the set of all microbial species and $Z_i\in\Zsc, i\geq 1$  be  a sequence  of distinct species.
 The model  does  not  constrain {\it  a  priori} the  number of species present in the  $J$ samples. The relative abundance of OTU $Z_i$ in sample $j$ is defined as
\begin{equation}\label{old.model}
P^j(\{Z_i\}) = \frac{\sigma_i\langle\bX_i,\bY_j\rangle_+^{2}}{\sum_{i'} \sigma_{i'}\langle\bX_{i'},\bY_j\rangle_+^{2}}
\end{equation}
where $\sigma_i\in (0,1)$, $\sigma_1>\sigma_2>\sigma_3>\ldots$, and $\bX_i,\bY_j\in \mathbb R^K$. The $k$-th components of $\bX_i$ and $\bY_j$ are denoted as $X_{k,i}$ and $Y_{k,j}$. We will explain the definitions of $\sigma_i$, $\bX_i$, $\bY_j$ and $K$ in the next paragraph. $\mathbb I(\cdot)$ is the indicator function and $x_+ = x\times \mathbb I(x>0)$. $\langle\cdot,\cdot\rangle$ denotes the standard inner product in $\mathbb R^K$. We define $Q_{i,j} = \langle \bX_i,\bY_j\rangle$. In addition, $\bsigma=(\sigma_i;i\geq 1)$, $\bX=(\bX_i;i\geq 1)$, $\bY=(\bY_j;j\leq J)$ and $\bQ = (Q_{i,j};i\geq 1, j\leq J)$.

We can interpret $\sigma_i>0$ as  a  summary  of  the  overall  abundance of species $i$ across samples. We call $\bX_i$ and $\bY_j$ species vector and sample vector, respectively. $\bX_i$ and $\bY_j$ are  latent  components  of the probability model. Differences  across  compositions $P^j$ are determined by the $\bY_j$ latent  vectors. Vectors $\bY_j$ can be interpreted as latent characteristics of the samples that affect their microbial compositions.
The model assumes that there are $K$  latent characteristics and  $\bX_i$ corresponds to the effects of these $K$ latent characteristics on the abundance of the species $Z_i$. 

The construction above implies that the angle $\phi_{j,j'}$ between $\bY_j$ and $\bY_{j'}$ determines the degree of similarity between compositions  $P^j$ and $P^{j'}$. 
Specifically, small $\phi_{j,j'}$ indicates that $P^j$ and $P^{j'}$ are  similar. When $\phi_{j,j'}=0$, compositions $P^j$ and $P^{j'}$ are  identical. 
Symmetrically, the angle $\varphi_{i,i'}$ between $\bX_i$ and $\bX_{i'}$ can be viewed as a measure of similarity between species $Z_i$ and $Z_{i'}$. When $\varphi_{i,i'}$ decreases towards zero, the correlation between $(P^j(\{Z_i\});j\leq J)$ and $(P^j(\{Z_{i'}\};j\leq J))$ increases to one.

The  prior specification in the model is as follows. First $\sigma_1>\sigma_2>\sigma_3\ldots$ are {\it  a  priori} ordered points from a Poisson process on $(0,1)$ with intensity $\nu(\sigma) = \alpha\sigma^{-1}(1-\sigma)^{-1/2}$. Second the $X_{k,i}$ random  variables  are  independent  Gaussian  $\Nsc(0,1)$,  $i=1,2,\ldots$, $k=1,2,\ldots,K$. We  can assume  for the  moment  that  the  $\bY_j$'s are  fixed.

The  resulting  marginal  prior distribution on  the  composition $P^j$ is a Dirichlet process \citep{boyuren}. In addition, $P^j$ and $P^{j'}$ are dependent for $j\neq j'$. To provide  some  intuition  on this construction of  Dirichlet process we  consider a similar model with $I<\infty$ species. For simplicity we set $\|\bY_j\|=1$, where $\|\cdot\|$ is the Euclidean norm of a real vector. 
The prior on $\bX_i$'s induces a standard normal distribution on $(Q_{1,j},\ldots,Q_{I,j})$. The prior distribution of $(Q_{i,j})_+^{2}$ is therefore a mixture of a point mass at zero and a $\text{Gamma}(1/2,1/2)$ distribution. Assume $(\sigma_1,\ldots,\sigma_I)$ are independent $\text{Beta}(\alpha/I, 1/2-\alpha/I)$ variables. It can be verified by moment generating function that the joint law of these ordered independent  Beta random variables converges to the law of a Poisson process on $(0,1)$ with intensity $\alpha\sigma^{-1}(1-\sigma)^{-1/2}$ when $I\to \infty$. The products $(\sigma_i(Q_{i,j})_+^{2},i=1,\ldots,I)$ then follow a mixture distribution of a point mass at zero and a $\text{Gamma}(\alpha/I,1/2)$. 
The normalized vector $(\sigma_i(Q_{i,j})_+^{2}/\sum_{i'}\sigma_{i'}(Q_{i',j})_+^{2},i=1,\ldots,I)$, conditioned on $(\mathbb I(Q_{1,j}>0), \ldots, \mathbb I(Q_{I,j}>0))$ follows a Dirichlet distribution with weights proportional to $\mathbb I(Q_{i,j}>0)$. If $I\to\infty$, we know that the $(\sigma_1,\ldots,\sigma_I)$ converges in distribution to the Poisson process with intensity $\nu$, and $(\sigma_i(Q_{i,j})_+^{2}/\sum_{i'}\sigma_{i'}(Q_{i',j})_+^{2},i=1,\ldots,I)$ becomes a Dirichlet process \citep{ferguson1973bayesian}. This holds also when $\|\bY_j\|\neq 1$ because the distribution of $(\sigma_i(Q_{i,j})_+^{2}/\sum_{i'}\sigma_{i'}(Q_{i',j})_+^{2},i=1,\ldots,I)$ does not depend on $\|\bY_j\|$.

For  inferential and visualization purposes  it  is  desirable  that  the  $\bY_{j}$  latent  vectors concentrate approximately on   a  low  dimensional  space. The resulting  $\bY_j$  are parsimonious latent factors that capture the variability of observed species abundances across samples. To this end, the model applies the prior  studied in \cite{dunsonfactor},
$$
\bY_j\sim \Nsc(\bzero,\text{diag}\{\gamma_1,\ldots,\gamma_K\}),
$$
where $\gamma_k$ rapidly decrease with $k$. The prior formalizes the  desiderata  of  having  the  norm  $\|\bY_j\|$    mostly   driven   by  the  first few components of $\bY_j$,  say  the  first  three components $(Y_{1,j},Y_{2,j},Y_{3,j})$,  and  the rest of the components, $(Y_{4,j},\ldots,Y_{K,j})$, vanish with  negligible  values. In different  words, only  a small  set  of $\bY_j$ entries---three in the example---are  relevant. This  approach is preferable  to  a  hyper-prior  on  the dimensionality of $\bY_j$ mainly  for computational convenience.

\subsection{Fixed effects}
\label{dp.fixed}
The goal of this subsection is to  model relationships between microbial compositions and samples' characteristics. For example, in studies of Inflammatory Bowel Disease (IBD) \citep{xochi2012,greenblum2012metagenomic,gevers2014treatment}, researchers were interested  in identifying  microbes that  correlate with the onset of IBD to develop therapeutic hypotheses. These  analyses   typically  utilize regression models  where  the  outcomes  coincide  with OTU abundances.
 Following  a similar strategy, we expand the model in Section \ref{prev.model}.  

Assume there are $L\geq 1$ observed covariates. We use $\bw_j=(w_{l,j};l=1,\ldots,L)$ to denote the covariates' values for sample $j$. The effects of this set of covariates on species $i$ are $\bv_i=(v_{l,i};l=1,\ldots,L)$. The collection of all $\bw_j$ and $\bv_i$ are $\bw=(\bw_1,\ldots,\bw_J)$ and $\bv = (\bv_1,\ldots,\bv_I)$. Our extended model directly modifies the random variables $Q_{i,j}$'s introduced in the definition of the model (\ref{old.model}), by adding a linear function of $\bw_j$ and an error term:
\begin{equation}
Q_{i,j} = \langle\bX_i, \bY_j\rangle + \langle\bv_i,\bw_j\rangle + \epsilon_{i,j},
\label{expand.model}
\end{equation}
where $\epsilon_{i,j}\overset{iid}{\sim} \Nsc(0,1)$ is the error term. Thus,
\begin{displaymath}
P^j(\{Z_i\}) = \frac{\sigma_{i} (Q_{i,j})_+^{2}}{\sum_{i'} \sigma_{i'}(Q_{i',j})_+^{2}}.
\end{displaymath}
 The inner product $\langle\bv_i,\bw_j\rangle$ represents the fixed effects of our model, whereas $\langle\bX_i,\bY_j\rangle$ represents the random effects. We fix the variance of the errors to one since the model for $P^j$ is invariant if we rescale all $Q_{i,j}$ variables by a fixed multiplicative term.

In this construction, $\bv_i$ and $\bw_j$ can be viewed as additional dimensions of $\bX_i$ and $\bY_j$ respectively. 
 The angle between $(\bw_j^{},\bY_j^{})^{}$ and $(\bw_{j'}^{},\bY_{j'}^{})^{}$, denoted as $\tilde\phi_{j,j'}$, measures the similarity between the microbial compositions $P^j$ and $P^{j'}$. As in model (\ref{old.model}), one can verify that the correlation  $\text{cor}(P^j(A),P^{j'}(A))$ is  monotone with respect to $\tilde\phi_{j,j'}$. Similarly, the angle between $(\bv_i,\bX_i)$  and  $(\bv_{i'},\bX_{i'})$, $\tilde\varphi_{i,i'}$, is representative of the correlation between abundances of species $i$ and $i'$ across samples. A small $\tilde\varphi_{i,i'}$ value makes the correlation between vectors $(P^j(\{Z_i\});j\leq J)$ and $(P^j(\{Z_{i'}\});j\leq J)$ close to one.

The  coefficients $\bv_i$ are \textit{a priori}  independent normal random variables with mean zero and variance one. 
When the latent factors $\bY$ are fixed, and the  prior for $\bX_i$ and $\sigma_i$ remains the same as in Section \ref{prev.model}, the microbial composition $P^j$, for  each $j=1,\ldots,J$, retains a marginal Dirichlet Process  distribution. More precisely, $P^j$  is  a Dirichlet process  with  concentration parameter $\alpha$. This can be shown using the same argument as in Section \ref{prev.model}.

\begin{figure}[htbp]
\centering
\includegraphics[scale=0.6]{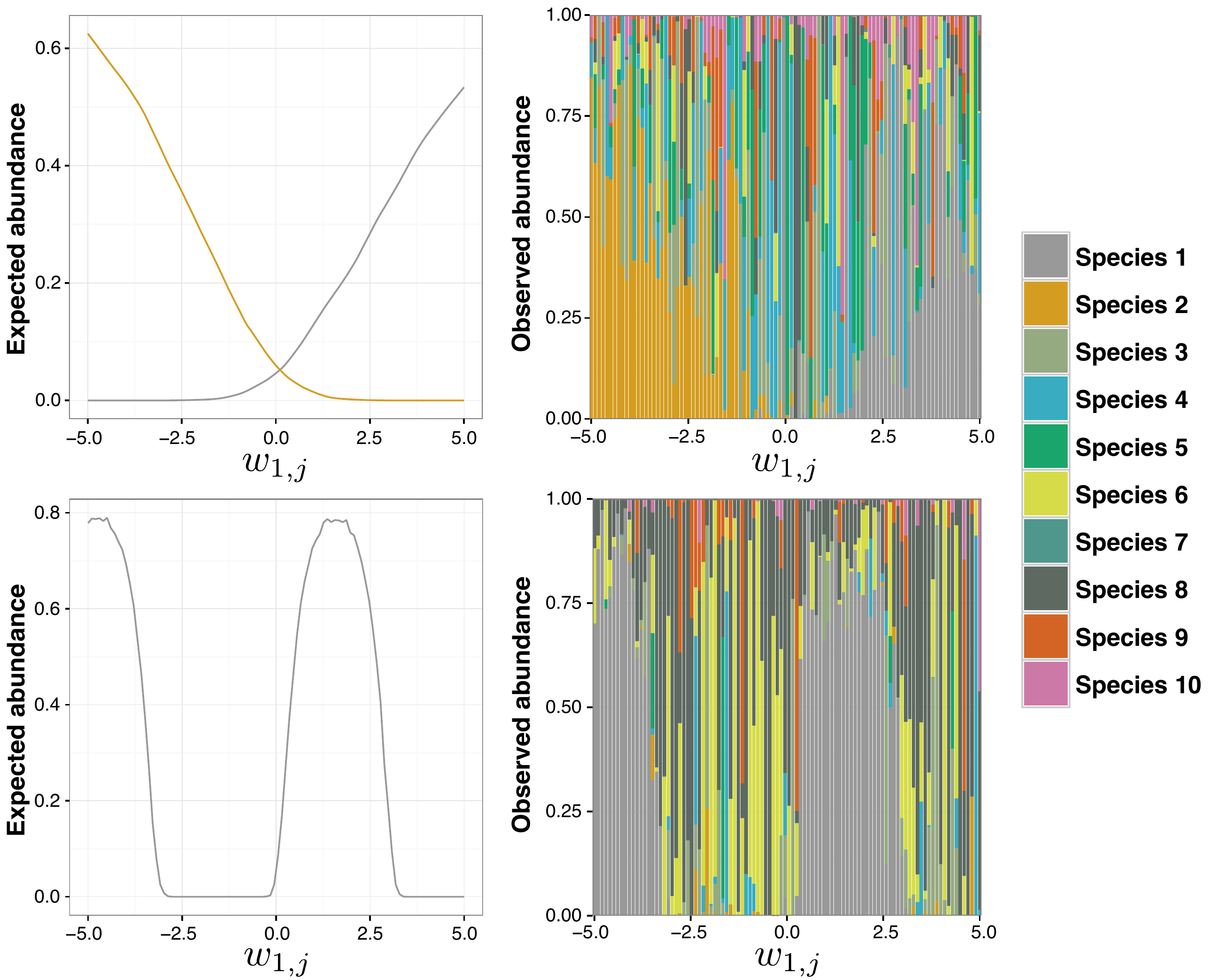}
\caption[Observed data generated from model in Chapter 2]{
Effect of a single covariate $w_{1,j}$ on microbial species abundances. 
We illustrate the expected abundances of species 1 and 2 when $w_{1,j}$ varies (\textbf{Left}) and the observed microbial abundances species 1-10 in one simulated dataset as $w_{1,j}$ changes (\textbf{Right}). 
We focus on a single sample $j$ and fix the random effects $\langle \bX_i,\bY_j\rangle$ in all simulations. 
Only the value of $w_{1,j}$ and the error terms $\epsilon_{i,j}$'s vary. 
The expected abundances are calculated by averaging over 1000 simulation replicates. 
We consider the cases where $Q_{i,j}=v_{1,i}w_{1,j}+\langle\bX_i,\bY_j\rangle+\epsilon_{i,j}$ with $v_{1,1}=5,$ $v_{1,2}=-5$ and $v_{1,i}=0$ for $i>2$ (\textbf{Top}) and 
$Q_{i,j}=v_{1,i}\sin(w_{1,j})+\langle\bX_i,\bY_j\rangle + \epsilon_{i,j}$ with $v_{1,1}=5$ and $v_{1,i}=0$ for $i>1$ (\textbf{Bottom}). 
The covariate $w_{1,j}$ varies from $-5$ to $5$ with $0.1$ increments.}
\label{fig1}
\end{figure}

It is  important not to misinterpret  the coefficients $\bv_i$. The species abundances are not linear functions of the covariates (see Figure \ref{fig1}). In certain cases, the relationship between the covariates and the species abundances is not monotone. Consider a single covariate $w_{1,j}$ and assume $Q_{i,j}=v_{1,i} w_{1,j} + \langle\bX_i,\bY_j\rangle +\epsilon_{i,j}$, where $v_{1,1}=5$, $v_{1,2}=1$ and $v_{1,i}=0$ when $i>2$. For simplicity, assume in addition $\sigma_i$ variables  all equal to $0.5$. 
When $w_{1,j}$ is  small, say $w_{1,j}\in (0,0.5)$, the abundances of species 1 and 2 increases with $w_{1,j}$. However, as $w_{1,j}$ gets larger, say $w_{1,j}>5$, species 1 will dominate all other species and the abundance of species 2 decreases to nearly zero.

\subsubsection{Models for data analysis}
\label{sub.spec.model}
In  our  analyses we considered longitudinal data with repeated measurements over time for each individual. 
Assume samples $j=1,\ldots,J$ are partitioned into $U$ groups,  i.e. $U$ distinct individuals. 
We use $u_j$ to identify the individual associated to sample $j$. We enforce the samples $j$ and $j'$ from the same individual $u$ ($u_j=u_{j'}=u$) to share common latent factors $\bY_u$. The \textit{longitudinal} version of model (\ref{expand.model})  utilizes 
\begin{equation}
Q_{i,j} = \langle \bX_i,\bY_{u_j}\rangle + \langle\bv_i,\bw_j\rangle + \epsilon_{i,j}.
\label{expand.model.long}
\end{equation}
The rationale for this model is that samples derived from the same individual tend to be similar. 
The covariates $\bw_j$ will   include  time  information  (e.g. individual's age)  for   each sample $j$. 
This version of the  model is tailored towards longitudinal data and studies with repeated measurements,  and it  allows one to visualize  time trends of microbial compositions. 

We will use a truncated version of model (\ref{expand.model}) or (\ref{expand.model.long}) in data analyses, which we call  the \textit{finite-species} model. Truncating the stick-breaking representation of Dirichlet process has been studied extensively in literature \citep{ishwaran2002exact}. The truncated process can be arbitrarily close to the Dirichlet process in total variation distance if the number of retained atoms is large. In our case, we truncate the model (\ref{expand.model.long}) at the number of observed species, $I$. This is sufficient for data analysis as the sequencing depth in microbiome studies is generally large enough to capture most of the microbial species of interest. With $I<\infty$ species the finite-species model is defined by
\begin{equation}
\begin{aligned}
Q_{i,j} &= \langle \bX_i, \bY_j \rangle + \langle\bv_i,\bw_j\rangle + \epsilon_{i,j}, & \qquad 
P^j(\{Z_i\})&= \frac{\sigma_i (Q_{i,j})_+^{2}}{\sum_{i'=1}^I \sigma_{i'} (Q_{i',j})_+^{2}},\;\;
\;
\;
\;
\;
\;
\;
i=1,\ldots,I.
\end{aligned}
\label{expand.model.finite}
\end{equation}
The prior for $\bX_i$ and $\bY_j$ remain identical. The prior for $\sigma_i$'s becomes
$
\sigma_i\overset{iid}{\sim}\text{Beta}(\alpha/I, 1/2-\alpha/I).
$

\subsection{Identifiability}
\label{sec.identify}

In this subsection we consider the identifiability of  the proposed model. 
Since the model is invariant under simultaneous rotations of  the  vectors $\bY_j$ and $\bX_i$, we cannot learn $\bY$ from the data. 
We discuss the identifiability of the correlation matrix $\bS$ associated to $\bSigma = \bY^\intercal \bY + \bI$, where $\bI$ is the $J\times J$ identity matrix.
 Similarly, since the  composition  $P^j$ is invariant to scale transformation of $\bsigma$ we will discuss identifiability  of the ratios $\sigma_i/\sigma_{i'}$  for $i\neq i'$. 
 We assume that the number of samples is finite and that covariates $\bw_j$'s are independent with $\mathbb E(\bw_j\bw_j^\intercal)$ of full rank.

We proceed assuming initially that  $(P^j(\{Z_i\});i\geq 1, j\leq J)$ are observable random variables. Recall that
\begin{equation}\label{eq:gauss}
(Q_{i,j};j\leq J)|\bv_i,\bY,\bw \sim \Nsc(\bw^\intercal\bv_i,\bSigma).
\end{equation}
Since we assume that  $P^j(\{Z_i\})$  is observable, we have that $P^j(\{Z_i\})=0$ implies $Q_{i,j}\leq 0$. 
Consider a set of new random variables, denoted as $(\Ptilde^j(\{Z_i\});i\leq I,j\leq J)$, where $\Ptilde^j(\{Z_i\})=\mathbb I(P^j(\{Z_i\})>0)$. From \eqref{eq:gauss}, the conditional distribution of $(\Ptilde^j(\{Z_i\});i\geq 1,j\leq J)$ given $(\bsigma,\bY, \bv, \bw)$ is
\begin{equation}
\begin{aligned}
&p(\Ptilde^j(\{Z_i\}),i\geq 1,\;\; j\leq J|\bsigma,\bY, \bv, \bw) \\
\propto& \prod_{i}\left[\int_{\Asc_{i}}(2\pi)^{-J/2}|\bSigma|^{-1/2}\times\exp\left(-\frac{1}{2}(\bQ_i-\bmu_i)^\intercal\bSigma^{-1}(\bQ_i-\bmu_i)\right)d\bQ_i\right].
\end{aligned}
\label{lik.fnc}
\end{equation}
Here $\bQ_i=(Q_{i,1},\ldots,Q_{i,J})$, $\bmu_{i} = \bw^\intercal\bv_i$, $\Asc_{i}=\bigtimes_{j=1}^J \Asc_{i,j}$ and $\Asc_{i,j}=(-\infty,0]$ if $\Ptilde^j(\{Z_i\})=0$,  while $\Asc_{i,j}=[0,\infty)$ when $\Ptilde^j(\{Z_i\})=1$. 
To illustrate the identifiability of the  parameters $(\sigma_i/\sigma_{i'};i\neq i'),\bS$ and $\bv$, we start with two simplified cases and then give a proposition.

\begin{enumerate}
\item {\it Without random effects} ($\bY = \bzero$). We first note that conditioning on $\bw$, for a fixed $i$, $(\Ptilde^j(\{Z_i\});j\leq J)$ are samples from a standard probit model \citep{albert1993bayesian}, where $\bv_i$ serves as  regression coefficients and the sample covariates are $\bw_j$. Based on the theory of generalized linear models $\bv_i$ is identifiable when $\mathbb E(\bw_j \bw_j^\intercal)$ is of full rank.

We then consider $(\sigma_i/\sigma_{i'};i\neq i')$. 
By construction, $$
\frac{P^j(\{Z_i\})}{P^j(\{Z_{i'}\})}=\frac{\sigma_i}{\sigma_{i'}}\frac{(Q_{i,j})_+^{2}}{(Q_{i',j})_+^{2}}.
$$
Here we use the convention that the ratio is zero whenever the denominator is zero. To ensure the identifiability of $(\sigma_i/\sigma_{i'};i\ge 1,j\leq J)$, we want to show  that if $$(P^j(\{Z_i\});i\ge 1, j\leq J),\bw|\bv,\bsigma\overset{d}{=} (P^j(\{Z_i\});i\ge 1, j\leq J),\bw|\bv',\bsigma',$$ then $\sigma_i/\sigma_{i'}=\sigma'_i/\sigma'_{i'}$ for all $i\neq i'$. Using the identifiability of $\bv_i$,
the  above  equality in distribution  implies  $\bv_i=\bv_i'$,  and  in turn the  equality  of  the  conditional distributions 
$p(((Q_{i,j})_+^{2},(Q_{i',j})_+^{2}),\bw_j|\bv,\bsigma)$  and $p(((Q_{i,j})_+^{2},(Q_{i',j})_+^{2}),\bw_j|\bv',\bsigma').$ 
This directly implies $\sigma_i/\sigma_{i'}=\sigma_i'/\sigma'_{i'}$ for all $i\neq i'$.

\item {\it Without fixed effects } ($\bv_i=\bzero$). We consider $\bsigma$ and $\bS$. The distribution of $(\Ptilde^j(\{Z_i\}),\Ptilde^{j'}(\{Z_i\}))$ is
$$
p(\Ptilde^j(\{Z_i\}),\Ptilde^{j'}(\{Z_i\})|\bsigma,\bY) = \frac{1}{2\pi} \int_{\Asc_{i,j}\times \Asc_{i,j'}} \!\!\!\!\!\!\!\!(1-S_{j,j'}^2)^{-1/2}\exp\left(-\frac{1}{2}\bq^\intercal\bS_{j:j'}^{-1}\bq\right)d\bq,
$$
where $S_{j,j'}$ is the correlation between $Q_{i,j}$ and $Q_{i,j'}$, and $\bS_{j:j'}$ is the correlation matrix of $(Q_{i,j},Q_{i,j'})$.  $\Asc_{i,j}=(-\infty,0]$ if $\Ptilde^j(\{Z_i\})=0$, while $\Asc_{i,j}=[0,\infty)$ if $\Ptilde^j(\{Z_i\})=1$. 
Corollary 3.12 in \cite{slepian} shows that $p(\Ptilde^j(\{Z_i\}),\Ptilde^{j'}(\{Z_i\})|\bv_i,\bY) $, when $\bv_i=\bzero$, is monotone with respect to $S_{j,j'}$. This implies that $S_{j,j'}$ is identifiable.

Using the same arguments as in the case where no random effect is present, one can show that the ratios $(\sigma_i/\sigma_{i'};i\neq i')$ remain identifiable.
\end{enumerate}

In the general case the identifiability of the model parameters, with both fixed and random effects,  is  described through Proposition 1 in Section S1 of the Supplementary Material.

\section{Posterior simulations and  visualization of covariates' effects}
\label{sec:3}
In this section we focus on posterior inference and computational aspects. 
In Section \ref{comp.1} we introduce an   algorithm for posterior simulations with the model  described  in Section \ref{dp.fixed}. Then, in Section \ref{comp.2} we propose graphical visualizations  to  illustrate    associations of microbial compositions  and covariates. These representations are  relevant for the analysis of microbial abundances   because, as we mentioned in Section \ref{dp.fixed}, a positive (or negative)   element  of  the  vector  $\bv_i$,  say  the  $l$-th  element,  does not  imply a  monotone  relation   between the  $l$-th   covariate  and the abundances of species $i$. To  illustrate the relation between the  $l$-th covariate and  species $i$,  we   estimate  how  the abundance  of species $i$ would  vary at hypothetical values of  the $l$-th covariate.

\subsection{Posterior simulations}
\label{comp.1}
We proceed with the finite-species model (\ref{expand.model.finite}). The likelihood  function is
$$
p(\bn|\bQ,\bsigma) \propto \left(\prod_{j=1}^J\prod_{i=1}^I(\sigma_i(Q_{i,j}))_+^{2})^{n_{i,j}}\right)\times\prod_{j=1}^J\left(\sum_{i=1}^I\sigma_i(Q_{i,j})_+^{2}\right)^{-n_j},
$$
and
\begin{equation}
\begin{aligned}
p(\bsigma,\bQ,\bX,\bY,\bv|\bn,\bw)\propto \left(\prod_{j=1}^J\prod_{i=1}^I(\sigma_i(Q_{i,j})_+^{2})^{n_{i,j}}\right)\times\prod_{j=1}^J\left(\sum_{i=1}^I\sigma_i(Q_{i,j})_+^{2}\right)^{-n_j}&\times \\
\pi(\bsigma,\bQ|\bX,\bY,\bv,\bw) \pi(\bX,\bY,\bv)&,
\end{aligned}
\label{post.den.raw}
\end{equation}
where $\pi$ indicates the prior. By introducing  positive latent random variables $\bT=(T_1,\ldots,T_J)$ as in \cite{james-latent}, we rewrite the  conditional  distribution,
\begin{equation}
\begin{aligned}
p(\bsigma,\bQ,\bX,\bY,\bv|\bn,\bw)\propto& \int \pi(\bsigma,\bQ,\bX,\bY,\bv|\bw)\times\\
&\prod_{j=1}^J\left\{\left(\prod_{i=1}^I(\sigma_i(Q_{i,j})_+^{2})^{n_{i,j}}\right)T_j^{n_j-1}\exp\left(-T_j\sum_i\sigma_i(Q_{i,j})_+^{2}\right)\right\}d\bT.
\end{aligned}
\label{post.density}
\end{equation}
We use a Gibbs sampler to perform posterior simulations. 
The algorithm iteratively samples $\bsigma,\bT,\bQ,\bX,\bY$ and $\bv$ from the full conditional  distributions.  We  describe the  two components of the algorithm.
\begin{enumerate}
\item  The first component samples $\bsigma,\bT$ and $\bQ$ from the  full conditional distributions. We note that $\sigma_1,\ldots,\sigma_I$,  given the remaining variables,  are conditionally   independent. The sampling   of  $(\sigma_1,\ldots,\sigma_I)$ from the full conditional  distribution is identical  as  in \cite{boyuren}. The  random variables  $T_1,\ldots,T_J$, given $(\bQ,\bn,\bsigma)$, are  conditionally  independent with Gamma distributions.  These random variables  can be straightforwardly generated from the full conditional distribution. To  complete this part of the  algorithm we  can  write
\begin{equation}
\begin{aligned}
p(Q_{i,j}|&\bn,\bQ_{-i,-j},\bT,\bsigma,\bX,\bY,\bw,\bv)\propto\\
&(Q_{i,j})_+^{2n_{i,j}}\times\exp(-T_j\sigma_i(Q_{i,j})_+^{2})\times\exp\left(-\frac{\left(Q_{i,j}-\langle\bX_i,\bY_j\rangle-\langle\bv_i,\bw_j\rangle\right)^2}{2}\right),
\end{aligned}
\label{post.Q}
\end{equation}
where $\bQ_{-i,-j}$ is identical to $\bQ$ with the only exception that it does not include $Q_{i,j}$. The density (\ref{post.Q}) indicates that  the $Q_{i,j}$'s random variables  are conditionally  independent. We also note  that the density in (\ref{post.Q}) is log-concave. We use these arguments to sample  $\bQ$ from the full  conditional distribution.

\item The second  component considers the sampling of $\bY,\bX$ and $\bv$ from the  full conditional distributions. Using expression (\ref{post.density}) we write
$$
p(\bX|\bn,\bsigma,\bT,\bQ,\bY,\bv,\bw)\propto \exp\left(-\sum_{i,j}\frac{(Q_{i,j}-\langle\bX_i,\bY_j\rangle-\langle\bv_i,\bw_j\rangle)^2}{2}\right)\times \pi(\bX).
$$
Recall   that  the  $\bX_i$'s  are  {\it  a  priori} independent normal  random variables. Therefore the full conditional  distribution of $\bX_i$ coincides with the conjugate posterior distribution  in a standard linear model \citep{lindley1972bayes}. Sampling  of  $\bY$ and $\bv$  from the  full conditional distributions  follows identical arguments. Indeed  the  prior  model  studied  in  \cite{dunsonfactor},  which  we  use for $\bY$, is conditionally conjugate.
\end{enumerate}

\subsection{Visualization of covariate effects}
\label{comp.2}
We consider  the partial derivatives
$$
\frac{\partial P^j(\{Z_i\})}{\partial w_{l,j}} := \partial\left[\frac{\sigma_i(\langle\bX_i,\bY_j\rangle + \langle\bv_i,\bw_j\rangle + \epsilon_{i,j})_+^{2}}{\sum_{i'} \sigma_{i'}(\langle\bX_{i'},\bY_j\rangle + \langle\bv_{i'},\bw_j\rangle + \epsilon_{i',j})_+^{2}}\right]\bigg/\partial w_{l,j}.
$$

The derivative $\partial P^j(\{Z_i\})/\partial w_{l,j}$ quantifies the abundance variation of species $i$ in sample $j$ in response to an infinitesimal increment of the $l$th component of $\bw_j$. We can estimate these derivatives from the data using the posterior approximation obtained by the algorithm in Section \ref{comp.1}. We use the estimates $\mathbb E\left(\partial P^j(\{Z_i\})/\partial w_{l,j}|\bn,\bw\right).$ For example, the top row of Figure \ref{dist_pw} summarizes the posterior distributions of $\partial P^j(\{Z_i\})/\partial w_{1,j}$, $j=1,\ldots,300$, for three species. Details on the  figure, including a  description of   the simulated data  that  generated  the panels,   are  provided in Section \ref{deriv.trend}. In species 1, the estimates of the derivatives are positive for the  majority  of  the  samples and tend to be large when $w_{1,j}>0$. We also note that the estimates of $\partial P^j(\{Z_i\})/\partial w_{l,j}$ are larger for samples in the subgroup $w_{2,j}=1$ than in the subgroup $w_{2,j}=0$. These results indicate that, for any $j=1,\ldots, 300$, if  we  could   increase (decrease) the value of $w_{1,j}$ while holding $w_{2,j}$ fixed, then one would expect an increase (decrease) of the  relative abundances of species 1, and this  trend  appears  more  pronounced in those samples with $w_{1,j}>0$ and $w_{2,j}=1$.

We also define
$$
P^j(\{Z_i\};\bw_0) := \frac{\sigma_i\left(\langle\bX_i,\bY_j\rangle + \langle\bv_i,\bw_0\rangle + \epsilon_{i,j}\right)_+^{2}}{\sum_{i'}\sigma_{i'}\left(\langle\bX_{i'},\bY_j\rangle + \langle\bv_{i'},\bw_0\rangle + \epsilon_{i',j}\right)_+^{2}};
$$
it is the abundance of species $i$ if  the covariates values of sample $j$ could  be enforced to be  equal to $\bw_0$.
 When estimating the effect of a binary covariate $w_{l,j}\in\{0,1\}$ on microbial compositions, we replace derivatives by differences:
\begin{equation}
\begin{gathered}
\frac{\Delta P^j(\{Z_i\})}{\Delta w_{l,j}}
:= P^j(\{Z_i\};\bw_{l,j}^{1})-P^j(\{Z_i\};\bw_{l,j}^{0}),
\end{gathered}
\label{discrete.effect}
\end{equation}
here $\bw_{l,j}^1$
 is identical to $\bw_j$ with the exception that the $l$-th component
 $w_{l,j}$ is set to be one  and symmetrically $\bw_{l,j}^0$ is specified with $w_{l,j}$ equal to zero. 
 Therefore $\Delta P^j(\{Z_i\})/\Delta w_{l,j}$ is  the variation of $P^j(\{Z_i\})$
 that one  would observe  by changing the value  of  a binary covariate.

We also consider the population-level associations between microbial compositions and a specific covariate, say the $l$-th covariate, when adjusting for all other covariates. To this end, we first define $\Pbar(\{Z_i\};\bw_0)$, the \textit{population average abundance} of species $i$ at a covariate value $\bw_0$, by
$$
\Pbar(\{Z_i\};\bw_0) := \frac{1}{J}\left(\sum_{j=1}^J P^j(\{Z_i\};\bw_0)\right),
$$
which quantifies the average abundance of species $i$ when all $J$ samples in the study have the same hypothetical  covariates values $\bw_0$. We estimate $\Pbar(\{Z_i\};\bw_0)$ from the data with $\mathbb E\left(\Pbar(\{Z_i\};\bw_0)|\bn,\bw\right)$.

To illustrate the association between the abundance of species $i$ and the $l$-th covariate, we visualize the variation of $\Pbar(\{Z_i\};\bw_0)$ as $w_{l,0}$
(the $l$th entry  of $\bw_0$)  varies and  all  other covariates remain  fixed at  $\bw_{-l,0}.$ 
This visualization is obtained by plotting the estimated $\Pbar(\{Z_i\};\bw_0)$ against $w_{l,0}$.
 We call the resulting curve  the \textit{population trend} of species $i$ with respect to the $l$-covariate at $\bw_{-l,0}$. 
In Figure \ref{dist_pw}, bottom row, we illustrate  population trends of three species with respect to the first covariate at $w_{2,0}=0$ and  at $w_{2,0}=1$.

Interactions  terms  for pairs of  covariates,  and  more  generally  functions of the  covariates, can be included in the proposed model. We specify a function $\mathbf f:\mathbb R^L\to\mathbb R^{L'}$
for interaction terms. One example is $\mathbf f(\bw_j)=w_{1,j}w_{2,j}$. The definition of $Q_{i,j}$ in (\ref{expand.model}) when interactions are incorporated becomes
$$
Q_{i,j} = \langle\bX_i,\bY_j\rangle + \langle\bv_i,(\bw_j,\mathbf f(\bw_j))\rangle + \epsilon_{i,j},
$$
where $\bv_i\in\mathbb R^{L+L'}$. In these  cases  variations of  the $l$-th coordinate of $\bw_j$   affect $\mathbf f(\bw_j)$ and translate into compositional  variations  equal to  $\partial P^j(\{Z_i\})/\partial w_{l,j}$ or  $\Delta P^j(\{Z_i\})/\Delta w_{l,j}$.

\section{Simulation study}
\label{sec:4}
In this section we focus on the model introduced in Section \ref{dp.fixed}, and we illustrate that 
 we can transform the model parameters into interpretable results on the relationship between covariates and microbial compositions. We also provide in  this section  a comparison between our model and a recent published latent factor model MIMIX \citep{MIMIX} which uses the logistic-normal distribution to link covariates to the relative abundances of species.  We illustrate in simulation scenarios that the proposed model has similar performance to the logistic-normal model even when the data is generated from MIMIX. When the degree of zero-inflation is large, our model tends to  outperform MIMIX  regression regardless of the underlying data generating model. The code for replicating the simulation studies is available from the Github repository \url{https://github.com/boyuren158/DirFactor-fix}.

In our simulation study we included  $I=100$  species  and  $J=300$ samples. The 300 samples are  taken  from $U=50$ individuals (see Section \ref{sub.spec.model}).  Each  individual  is  measured six times. The read-depth of each sample  is  $n_j=10^{5}$. We simulate $\bsigma$  using independent  Beta densities  with mean $0.2$  and variance $0.1$. As we discussed in Section \ref{prev.model}, $\sigma_i$ represents the average abundance of species $i$ across all samples.  We included in the simulation a continuous covariate $w_{1,j}$, generated  from independent $\Nsc(0,1)$ distributions, and a binary covariate $w_{2,j}$, generated from independent $\text{Bernoulli}(0.5)$. We also use the interaction term $w_{1,j}\times w_{2,j}$ to specify scenarios where effects of $w_{1,j}$ differ in  the groups $w_{2,j}=0$ and $w_{2,j}=1$. We  will later  discuss  in Section \ref{application}  this type of interaction in a microbiome study for type 1 diabetes.
 
For the latent factors $\bY$ we assumed $\bY_u\in\mathbb R^4$. For the first half of the individuals, $u=1,\ldots,25$, we set $Y_{3,u}=Y_{4,u}=0$ while for the other half, $u=26,\ldots,50$, we set symmetrically $Y_{1,u}=Y_{2,u}=0$. The non-zero components in $\bY_u$ were simulated  independently from a $\Nsc(0,1)$ density. This specification of $\bY$ makes the correlation matrix $\bS$  block diagonal (see  Figure \ref{control.corr}(b)).

We  simulate  the first eight species  with positive $v_{1,i}$'s and the following eight species ($i=9,\ldots,16$) with negative $v_{1,i}$'s. As detailed  in Table \ref{v.spec} the first 16 species  abundances  correlate  with  $w_{2,j}$. Moreover, we make the assumption that  some  of the  trends   with  respect to $w_{1,j}$ are either amplified or reversed when we  contrast  the  two   groups  $w_{2,j}=1$ and $w_{2,j}=0$. All other species ($i>16$) have the corresponding $\bv_i$  coefficients equal to $\bzero$ (Table \ref{v.spec}).
\begin{table}
\centering
\small
\caption{Specification of $\bv$ in the simulation study.}
$$
\left[\begin{tabular}{c|ccccccccccccccccccc}
Species $(i)$&1&2&3&4&5&6&7&8&9&10&11&12&13&14&15&16&17&\ldots&100\\
\hline
$v_{1,i}$(for $w_{1,j}$)&5&5&5&5&5&5&5&5&-5&-5&-5&-5&-5&-5&-5&-5&0&\ldots&0\\
$v_{2,i}$(for $w_{2,j}$)&5&5&5&5&-5&-5&-5&-5&5&5&5&5&-5&-5&-5&-5&0&\ldots&0\\
$v_{3,i}$(for $w_{1,j}\cdot w_{2,j}$)&10&-5&-5&-10&10&-5&-5&-10&-10&5&5&10&-10&5&5&10&0&\ldots&0
\end{tabular}\right]
$$
\label{v.spec}
\end{table}

We further examine the robustness of our method by checking its performances when the link function between $P^j$ and $Q_{i,j}$ is misspecified.  In particular, we apply our method to data simulated using the following specification of $(P^j(\{Z_i\});i\leq I, j\leq J),$
\begin{equation}
P^j(\{Z_i\}) = \frac{\sigma_i Q_{i,j}^{+}}{\sum_{i'}\sigma_{i'} Q_{i',j}^{+}}.
\label{expand.model.mis}
\end{equation}
The specification of $\bsigma$, $\bv$, $\bY$ and $\bw$ remains the same as described in the previous paragraphs.

\subsection{Estimating species and samples parameters $\bv$ and $\bS$} 
\label{bv.bS.sec}
We first consider estimation of $\bv$ and $\bS$ between individuals when the model is correctly specified. Recall,  from Proposition 1 in the Supplementary Material, $\bv$ is identifiable when $\text{trace}(\bSigma)$ is assumed  fixed at a constant value. We  assume $\text{trace}(\bSigma)=1$ and compute the posterior distribution of $\bv/\sqrt{\text{trace}(\bSigma)}$. 
The performance of the estimate of $\bS$ is measured by the RV-coefficient \citep{rvcoef}, which is bounded between zero and one, between the estimated $\bS$ and the actual value of $\bS$.  An  RV-coefficient close to one indicates that the estimate is close to the  parameter $\bS$ used in simulations.

\begin{figure}[htbp]
\centering
\includegraphics[scale=0.645]{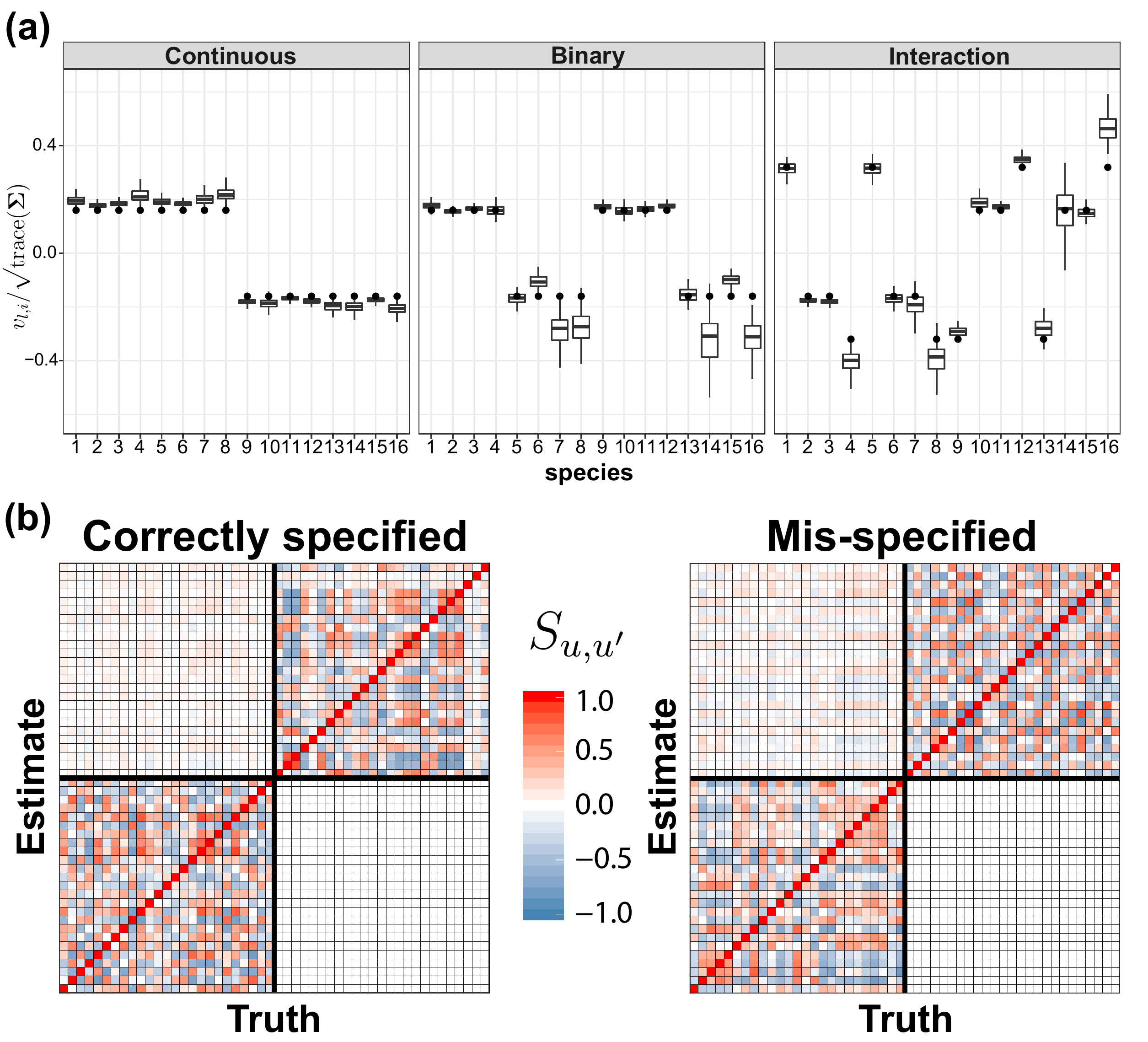}
\caption{Estimates of $\bv_i$, $i=1,\ldots,16$, and $\bS$ between individuals. (\textbf{a}) Posterior distributions of $\bv_i/\sqrt{\text{trace}(\bSigma)}$, $i=1,\ldots,16$. The posterior distributions are visualized by boxplots. The corresponding  values of $\bv_i/\sqrt{\text{trace}(\bSigma)}$  used for data simulation are indicated by  dots. (\textbf{b}) Posterior mean of the correlation matrix $\bS$ between individuals (values  above  the main diagonal) compared to the truth  (values  below  the main diagonal) in one simulation when the model is correctly specified (\textbf{Left}) and misspecified (\textbf{Right}).}
\label{control.corr}
\end{figure}

In Figure \ref{control.corr}(a) we illustrate the estimates of $\bv_i$, $i=1,\ldots,16$, in one simulation. The posterior means of  $\bv_i$ for the first 16 species are in general close to the corresponding  values  of the  simulation scenario. 
One exception is species 16, whose average relative abundance is the lowest ($8.1\times 10^{-5}$) among the first 16 species.
 In the left panel of Figure \ref{control.corr}(b), we illustrate the posterior mean of $\bS$ between individuals in one simulation. The estimate is close to the actual value of $\bS$ with an RV coefficient between them equal to 0.98, although the estimate indicates weak correlation between two independent subgroups (subject 1-25 and subject 26-50).

When the model is misspecified (see equation (\ref{expand.model.mis})), the estimates of $\bv$ are not comparable to the corresponding  values of the simulation scenarios. However, this result does not discourage the application of model (\ref{expand.model}) when estimating effects of covariates on microbial compositions.  The model can still capture the derivatives and population trends (see Section \ref{deriv.trend}), which directly describe the covariates' effects. The estimate of $\bS$, on the other hand, is only  minimally  affected by model misspecification and preserves its closeness to the actual value of $\bS$ (Figure \ref{control.corr}(b), right panel). The RV coefficient between the estimate and the actual value of $\bS$ is 0.96 in this case.

We then repeat  the simulation for 50 times under the correctly specified model as well as the misspecified model to verify  the  observed  accuracy levels. We fix across  all simulation replicates the values of $\bsigma$. When the model is correctly specified, the mean squared errors (MSEs) between the posterior means of rescaled $\bv_i$ coefficients  and the   values of the simulation scenario across the 50 simulation replicates   are  comparable across species. The smallest average MSE across 50 replicates is $4.3\times 10^{-6}$ for species 18 (the standard deviation of the estimate is $5.8\times 10^{-6}$) and the largest average MSE is $5.2\times 10^{-3}$ for species 14 (the standard deviation of the estimate is $2.3\times 10^{-2}$). The RV coefficients between the posterior means of $\bS$ and the actual value of $\bS$ across 50 replicates are close to one, whether the model is correctly specified or not. When the model is correctly specified, the mean and the standard deviation of the RV-coefficients are 0.964 and 0.009. When the model is misspecified, the mean and the standard deviation are 0.960 and 0.012. We diagnosed the mixing of the MCMC sampler for our model with $\hat R$ statistics \citep{Rhat}. The $\hat R$ statistics indicate that when the model is correctly specified, mixing is reached for rescaled parameter $v_{l,j}$ and eigenvalues of $\bS$ after 60,000 iteration. See Section S2 of the Supplementary Material for details.

The Bayesian  model  can be embedded into  a  permutation  procedure to detect whether a covariate $w_{l,j}$ is associated with the microbial composition or not. The  null and alternative hypotheses that  we consider are
$
H_0: \bv_{l}=\bzero_I\;\; vs.\;\; H_A: \bv_l\neq \bzero_I,
$
where $\bzero_I$ is a vector  of  zeros.
 We permute covariate values  $w_{l,j}$ across  samples and estimate, under  $H_0$, the distribution of $\|\hat\bv_l\|$, where $\hat\bv_l$ is the posterior mean of $\bv_l$.
     Permutation  is  one  possible  approach  to     estimate    the   $\|\hat\bv_l\|$  distribution  under   $H_0$,  which is applicable  if  covariates are   independent  or  nearly  independent.
  We  finally  compare  the  actual $\|\hat\bv_l\|$ value, with the  estimated distribution  (see  Section S3 of the Supplementary Material for an example).  
  One could  apply other   approaches  to  generate in silico  datasets  under $H_0$.  For  example, the parametric bootstrap  can replace the observed $w_{l,j}$ values with  samples from  estimates of the  conditional distributions $p(w_{l,j}|w_{-l,j})$,  where $w_{-l,j}$ indicates  the values of all covariates  except  $w_{l,j}$.

\subsection{Visualizing the relationship between covariates and  microbial compositions}
\label{deriv.trend}
As we mentioned in Section \ref{dp.fixed}, the values of $\bv$  do not directly  express  the sign and the magnitude  of  the covariates' effects on microbial compositions. Recall for  example  that  a  positive  $v_{l,i}$  might  correspond  to  a  decreasing trend with respect to the covariate $w_{l,j}$. 
This can happen when the $v_{l,i'}$ of another species $i'$ is  larger than $v_{l,i}$. 
The  goal  of  this subsection  is  to evaluate if we can estimate responses of species abundances to variations  of covariates of interest.

\begin{figure}
\centering
\includegraphics[scale=0.4]{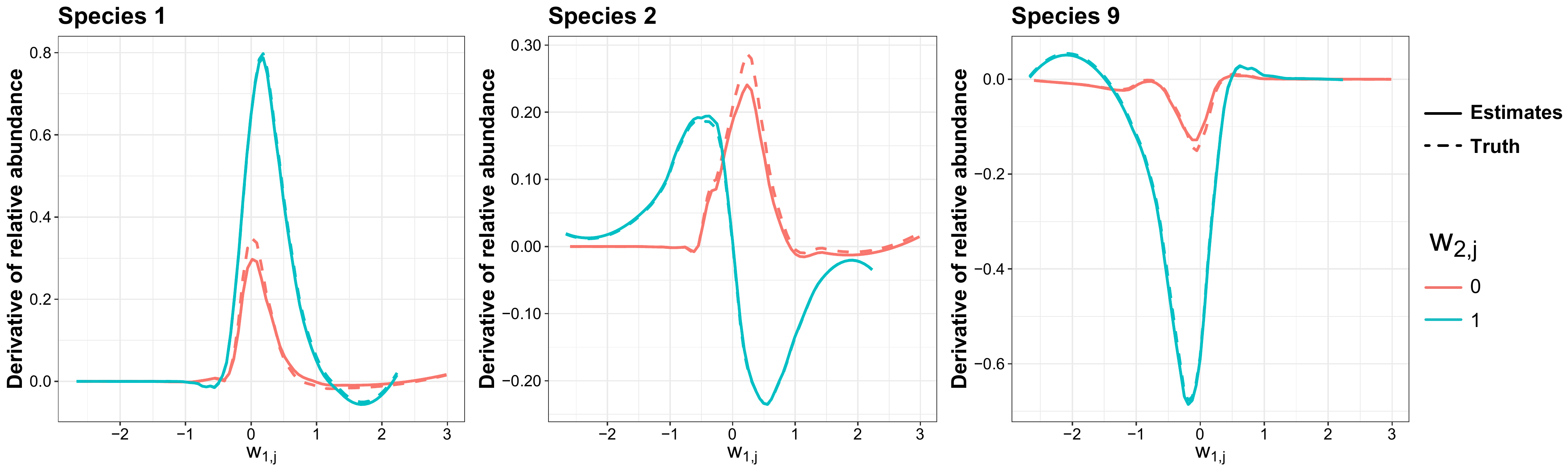}
\includegraphics[scale=0.4]{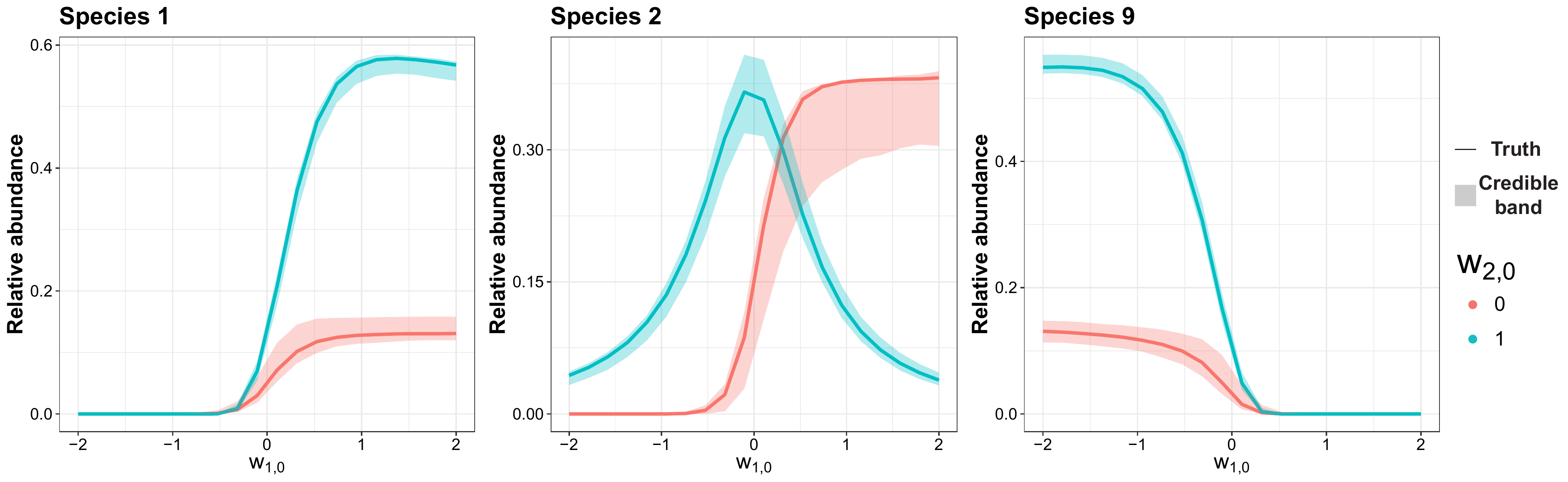}
\caption{Posterior estimates of individual-level and population-level relationship between covariate $l=1$ and relative abundances when the model is correctly specified. 
(\textbf{Left}) Increasing trend for the group $w_{2,j}=0$ and the group $w_{2,j}=1$. (\textbf{Middle}) Increasing trend for the group $w_{2,j}=0$ and  non-monotone trend for the group $w_{2,j}=1$. (\textbf{Right}) Decreasing trend for the group $w_{2,j}=0$ and the group $w_{2,j}=1$. 
Each  curve in the \textbf{top}   panels  is generated 
 by  computing  the individual  posterior estimates of $\partial P^j(\{Z_i\})/\partial w_{l,j}$, for each sample $j$,  which  then  become  the  input  of a  LOWESS procedure. 
The  \textbf{bottom} panels  illustrate the posterior distribution of the population trends.}
\label{dist_pw}
\end{figure}

We consider the visualization approaches described in Section \ref{comp.2}. We first focus on the estimates of the derivatives $\partial P^j(\{Z_i\})/\partial w_{l,j}$.
 These provide, for each individual sample,  estimates  of   the  variation in microbial  abundance  resulting from  an  infinitesimal  increment  of  a  specific  covariate $w_{l,j}$, while the other covariates  remain fixed. 
 The results for three representative species are summarized in the top panels of Figure \ref{dist_pw}. 
 The X-axes indicate the value of $w_{l,j}$ and the Y-axes the value of $\partial P^j(\{Z_i\})/\partial w_{l,j}$. 
 Each solid curve in these figures is generated 
 by  computing  the posterior means of $\partial P^j(\{Z_i\})/\partial w_{l,j}$, for each sample $j$,  which  then  become  the  input  of a  LOWESS algorithm. 
 We also calculate the actual values of $\partial P^j(\{Z_i\})/\partial w_{l,j}$ using the $\bsigma,\bX,\bY$ and $\bv$  parameters that generated the data. 
 The actual values of the partial derivatives are visualized with dash lines. 
 In Section S4 of the Supplementary Material, we also plot the posterior mean of $\partial P^j(\{Z_i\})/\partial w_{l,j}$ versus the actual value of $\partial P^j(\{Z_i\})/\partial w_{l,j}$ for each sample $j$, along with the 95\% credible intervals for $\partial P^j(\{Z_i\})/\partial w_{l,j}$.

We then focus on the population level estimates of covariates' effects by visualizing the population trend of species $i$ with respect to a given covariate (see Section \ref{comp.2}). Population trends of three representative species with respect to $w_{1,0}$ at different values of $w_{2,0}$ are summarized in Figure \ref{dist_pw}, bottom panels. 
The X-axes  indicate the value of $w_{1,0}$ and the Y-axes the population average abundance $\Pbar(\{Z_i\};\bw_0).$ The shaded areas indicate the pointwise 95\% credible bands of population trends. 

\begin{figure}
\centering
\includegraphics[scale=0.4]{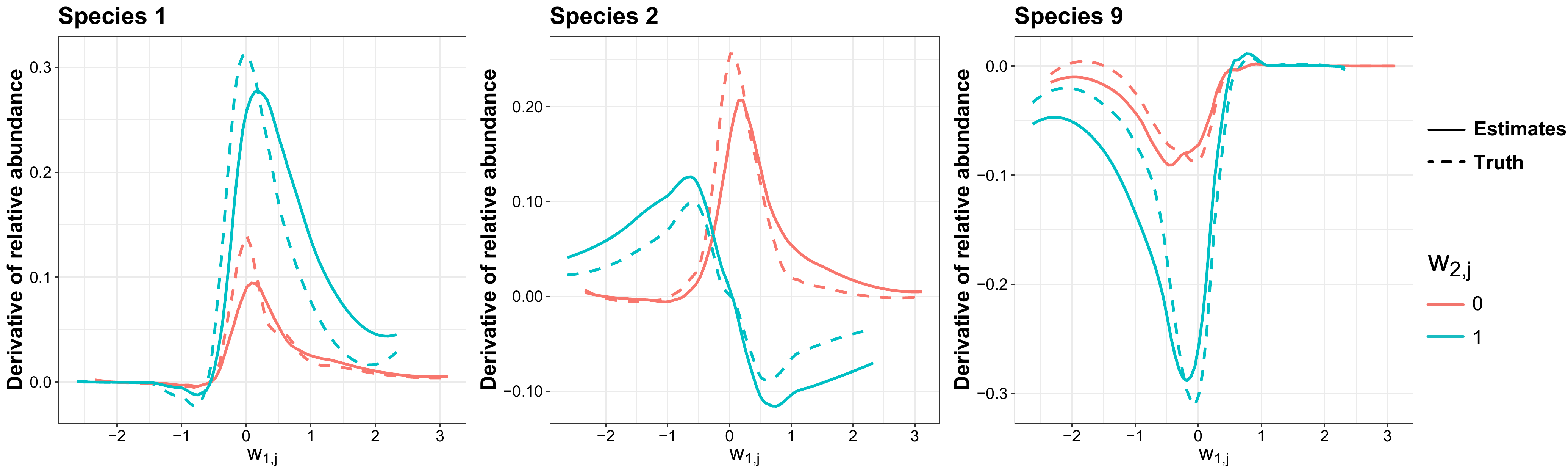}
\includegraphics[scale=0.4]{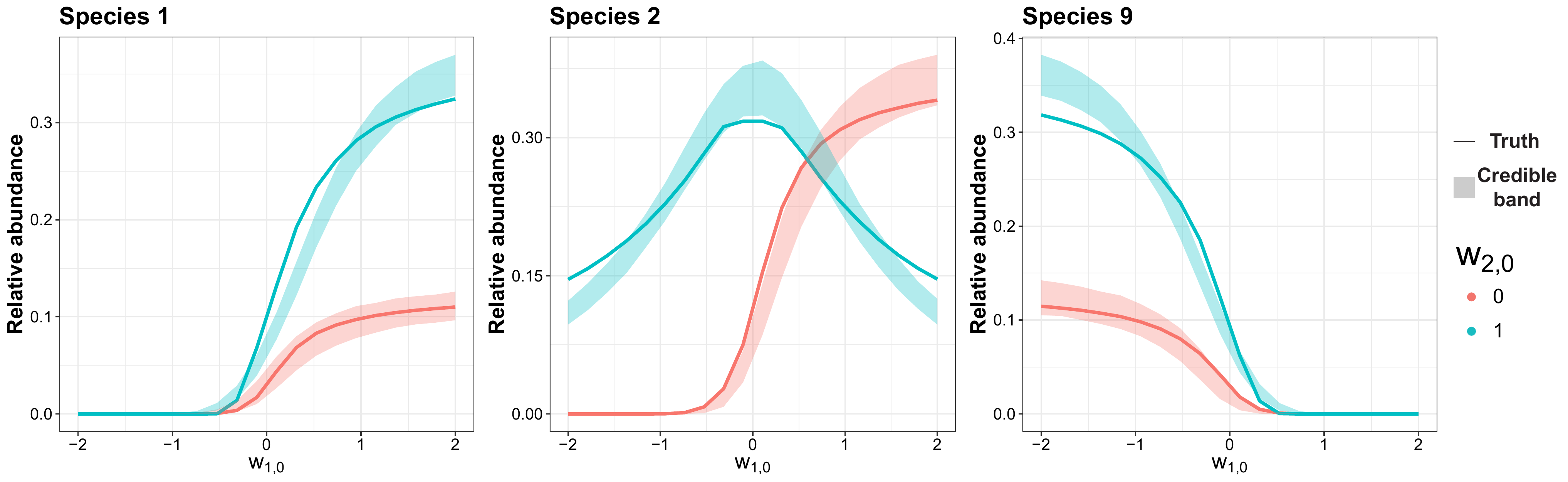}
\caption{Posterior estimates of individual-level and population-level relationship between covariate $l=1$ and relative abundances when the model is misspecified. }
\label{dist_pw_mis}
\end{figure}

When the model is misspecified, the comparisons of estimated derivatives and population trends to the truth are shown in Figure \ref{dist_pw_mis}. 
To compute the actual derivatives and population trends, we use the specification of $P^j(\{Z_i\})$ in (\ref{expand.model.mis}). From the top panels of Figure \ref{dist_pw_mis}, we observe that the estimates of the derivatives capture the sign of the actual values. However, the estimates are not as close to the actual values of the derivatives as in the case where the model is correctly specified. 
 This result is expected as we erroneously assume that $P^{j}(\{Z_i\})$ depends on $(Q_{i,j})_+^{2}$ instead of $(Q_{i,j})_+$. Bottom panels of Figure \ref{dist_pw_mis} illustrate that the estimated population trends follow the actual trends,  but the posterior credible bands do not cover the truth as  in the  previous example, where the model is correctly specified.

We repeat the simulation for 50 times under the correctly specified model as well as under the misspecified model.
  For each species $i$, we use MSE between the posterior mean of $(P^j(\{Z_i\});j\leq J)$ derivatives and the corresponding values of our simulation model. 
  In Supplementary Figure S5.1 (top panel), we plot the distributions of MSEs across simulation replicates. 
  This figure confirms the results in Figure \ref{dist_pw} and Figure \ref{dist_pw_mis}.
   The estimates of derivatives in the correctly specified model are  closer to the truth  compared to the estimates with the misspecified model. 
   For both, correctly specified and misspecified models, 
   the mean MSE across  replicates reaches its maximum for species 9, with value $8.7\times 10^{-4}$ under the correctly specified model and $3.8\times 10^{-2}$ under the misspecified model. 

We then consider the estimates of population trends in the 50 replicates. In Supplementary Figure S5.1 middle and bottom panels, we illustrate the estimated population trends in three species, when the model is correctly specified and misspecified. When the model is misspecified, the overall shape of each band still mirrors the actual trend,  but  the confidence  band does  not   cover the actual trend in  a  few intervals of $w_{1,0}$. 

\subsection{The Logistic-normal model}
\label{dir.mimix.comp}
We conclude the simulation study with a comparison of our model (referred to as DirFactor) to MIMIX \citep{MIMIX}, a logistic-normal model with latent factors. MIMIX employs a low-dimensional latent structure that is shared by both the fixed effects and the random effects to highlight the relationships between microbial species. The major difference between MIMIX and our model lies in the distribution assumption for $P^j$. In MIMIX, the distribution of $P^j$ follows a logistic-normal distribution:
\begin{equation}
P^j(\{Z_i\}) = \frac{\exp(Q_{i,j})}{\sum_{i'}\exp(Q_{i',j})}.
\label{MIMIX.link}
\end{equation}
A  characteristic of this specification is that the relative abundances of species are strictly positive and not tailored to zero-inflated microbiome data. By contrast, our specification of $P^j$ assigns non-zero mass to zero, which means that our model allows for explicit zero-inflation. In this subsection, we are interested in comparing the estimation performance of our model to that of MIMIX.

We focus on the accuracy of the estimated population trends for the continuous covariates $w_{1,j}$. The accuracy is evaluated in two aspects: root mean-squared errors (RMSE) of the estimated population trends and coverages of the estimated credible bands of the population trends. The first metric is a universal summary of the bias and variance of the estimates while the second metric is  used to  evaluate the reported  uncertainty on the estimates. We generate two sets of simulation datasets. The first set of data is generated using the link function (\ref{old.model}) in our model whereas the second set uses the link function (\ref{MIMIX.link}) in MIMIX. The specifications of $\bv, \bw, \bX, \bY$ are the same for both sets and are described at the beginning of Section \ref{sec:4}.

For each simulated dataset, we impose additional zero-inflation via hard truncation of $P^j$ at $10^{-3}$ and $10^{-2}$. The larger the threshold the higher the degree of zero-inflation introduced in the simulated dataset. We also examine the effect of overdispersion. Specifically, for fixed $\bv, \bw, \bX, \bY$, we generate three datasets based on them with $\text{var}(\epsilon_{i,j})=1$, $\text{var}(\epsilon_{i,j})=5$ and $\text{var}(\epsilon_{i,j})=10$ to represent low overdispersion, medium overdispersion and high overdispersion. We finally consider the effect of overdispersion in the distribution of read depths $n_j$. Once relative abundances $(P^j(\{Z_i\});j\leq J, i\leq I)$ are simulated, we generate the OTU counts with three different distributions of $n_j$: a Poisson distribution with mean $10^5$, a negative Binomial distribution with mean $10^5$ and variance $10^9$ (moderate overdispersion) and a negative Binomial distribution with mean $10^5$ and variance $4\times 10^{10}$ (large overdispersion).

We use a B-spline basis of $w_{1,j}$  
 both 
 when  we  produce  inference  based on 
 our model or  MIMIX  in the simulation  study  (i.e.  we  don't  directly  incorporate  the  $w_{1,j}$  values   within the  models). This adds  flexibility  in the relation  between covariates  and microbial compositions. We  recommend the use  of  splines  or  other transformations 
   when the number of covariates is considerably  lower than the number of samples, as  in our simulation study. The B-spline basis we used is of degree three with  internal knots at -1, 0 and 1 and two boundary knots at -2 and 2. We simulate 50 instances of $\bv,\bw,\bX$ and $\bY$. For each simulation replicate of $\bv,\bw,\bX$ and $\bY$, we generate  datasets based on combinations of  different link function (\ref{old.model}) and (\ref{MIMIX.link}), three different truncation levels, three overdispersion levels and three distributions of $n_j$.

In each simulation replicate, we estimate the population average abundance (see Section \ref{comp.2} for its definition) of each species at 20 different values of $w_{1,0}$ equally spaced between -2 and 2. We report the average RMSE between the resulting vector of  estimates  and   simulation scenario parameters  across all species and 50 simulation replicates as well as two values of $w_{2,0}$. We also report the  coverage of  the 95\% credible intervals of the population trends for $w_{1,0}\in(-2,2)$ in the 50 replicates averaging across all species and two values of $w_{2,0}=0,1$.
For $n_j$   generated   from the Poisson distribution, the RMSEs are shown in Table \ref{sim.compare.mse} and the coverage probabilities are included in Table \ref{sim.compare.ci}. For  the other two  distributions of the $n_j$ counts we illustrate the results in Section S6 of the Supplementary Material.

\begin{table}
\centering
\begin{tabular}{c|ccc|ccc|ccc|ccc|}
      & \multicolumn{6}{c|}{Simulated from DirFactor} & \multicolumn{6}{c|}{Simulated from MIMIX} \bigstrut[b]\\
\cline{2-13}      & \multicolumn{3}{c|}{DirFactor} & \multicolumn{3}{c|}{MIMIX} & \multicolumn{3}{c|}{DirFactor} & \multicolumn{3}{c|}{MIMIX} \bigstrut\\
\hline
Threshold     & 0 & $10^{-3}$ & $10^{-2}$ & 0 & $10^{-3}$ & $10^{-2}$ & 0 & $10^{-3}$ & $10^{-2}$ & 0 & $10^{-3}$ & $10^{-2}$ \bigstrut[t]\\
\hline
 
$\var(\epsilon_{i,j})=1$ &1.0&1.2&1.4&15.3&24.3&44.8&1.3&2.5&3.8&1.3&1.5&1.8\\
$\var(\epsilon_{i,j})=5$ &3.5&3.5&3.4&35.4&65.9&74.1&5.4&7.9&8.8&1.4&2.9&4.1\\
$\var(\epsilon_{i,j})=10$ &4.5&4.6&4.7&66.0&96.6&154.3&10.6&11.3&11.7&1.7&3&4.7\bigstrut[b]\\
\hline
\end{tabular}%
\caption{Average RMSE of estimated population mean abundances at 20 different values of $w_{1,0}$ equally spaced between -2 and 2 across simulation replicates for our model (DirFactor) and MIMIX. The threshold parameter indicates at which value we truncate the simulated $P^j(\{Z_i\})$'s to zero. We consider two scenarios where the data is generated from DirFactor and MIMIX respectively. The read depths are generated from a Poisson distribution with mean $10^5$. All RMSEs in the table are multiplied by $10^3$.}
\label{sim.compare.mse}
\end{table}

\begin{table}
\centering
\begin{tabular}{c|ccc|ccc|ccc|ccc|}
      & \multicolumn{6}{c|}{Simulated from DirFactor} & \multicolumn{6}{c|}{Simulated from MIMIX} \bigstrut[b]\\
\cline{2-13}      & \multicolumn{3}{c|}{DirFactor} & \multicolumn{3}{c|}{MIMIX} & \multicolumn{3}{c|}{DirFactor} & \multicolumn{3}{c|}{MIMIX} \bigstrut\\
\hline
Threshold     & 0 & $10^{-3}$ & $10^{-2}$ & 0 & $10^{-3}$ & $10^{-2}$ & 0 & $10^{-3}$ & $10^{-2}$ & 0 & $10^{-3}$ & $10^{-2}$ \bigstrut[t]\\
\hline
$\var(\epsilon_{i,j})=1$ &0.99&0.97&0.95&0.92&0.86&0.80&0.95&0.89&0.89&1.00&0.93&0.89\\
$\var(\epsilon_{i,j})=5$ &0.97&0.97&0.95&0.94&0.87&0.80&0.96&0.91&0.90&1.00&0.95&0.90\\
$\var(\epsilon_{i,j})=10$ &0.94&0.91&0.90&0.94&0.86&0.77&0.90&0.88&0.83&0.94&0.90&0.84\bigstrut[b]\\
\hline
\end{tabular}
\caption{Coverage of the posterior distribution of the  population trend (defined in Sec 3.2). We average across  species and across  $w_{1,0}$ values between -2 and  2  and  $w_{2,0}=0$ and $w_{2,0}=1$.  The threshold parameter indicates at which value we truncate the simulated $P^j(\{Z_i\})$'s to zero. The coverage is calculated using simulation replicates. The read depths are generated from a Poisson distribution with mean $10^5$. We consider two scenarios where the data is generated from DirFactor and MIMIX respectively.}
 \label{sim.compare.ci}
\end{table}

From the results we find that the  proposed  DirFactor model  shows  little  sensitivity  to the degree of zero-inflation. 
Setting $P^j(\{Z_i\})$ to be zero when its value is below a given threshold does not affect  accuracy. On the other hand, when the threshold for truncating $P^j(\{Z_i\})$ increases, the  accuracy of MIMIX  tends to decrease.  The RMSE of MIMIX  increases  with  this  threshold parameter,  regardless of the data generating models (\ref{old.model}) and (\ref{MIMIX.link}) and the level of overdispersion $\text{var}(\epsilon_{i,j})$.  The  performances in terms  of coverage  of the two models  appear  comparable  even when $\text{var}(\epsilon_{i,j})$ is large. These findings are confirmed when the distribution of $n_j$ counts is a negative binomial distribution. 
But prediction accuracy and coverage of the two models decrease significantly when the overdispersion of the negative binomial distribution is large (mean $=10^5$ and variance $=4\times10^{10}$).
See Supplementary Tables S6.3 and S6.4 for details.

We conclude this subsection with a posterior predictive procedure  to evaluate  and  compare Bayesian models.
   For distinct  Bayesian models, we compute  leave-one-out 95\% posterior predictive intervals of the relative abundance ($P^j(\{Z_i\})$)  of a species $i$ in sample $j$ using   the  available data, with  sample $j$ excluded. 
   The predictive intervals are generated using Pareto smoothed importance sampling \citep{Vehtari1,Vehtari2}. 
   We calculate the predictive intervals for all samples and all species in the data. We then derive the proportion of samples whose observed relative abundances $n_{i,j}/n_j$ of species $i$ are covered by the corresponding leave-one-out predictive intervals. We define the mean coverage probability of the model  by   averaging  these proportions across  species. In Section S7 of the Supplementary Material, we illustrate the approach in the  comparison of our Bayesian model and MIMIX \citep{MIMIX}. Limitations of leave-one-out cross-validation  in terms of stability have been previously discussed (Kohavi, 1995), the use of  the  procedure in our work serves the main purpose  of  producing interpretable  summaries that  integrate our evaluations  and comparisons of  regression methods.

\section{Microbiome analyses for type 1 diabetes in early infancy}
\label{application}
We use the longitudinal model in Section \ref{sub.spec.model} to  evaluate associations between gut microbiome  compositions, clinical variables  and demographic characteristics of infants in the DIABIMMUNE project \cite{tommi}. 
The DIABIMMUNE project  collected  longitudinal microbiome data in  157 infants over a period  up  to 1600 days after birth. 
 Infants  were  enrolled from Finland, Estonia and Russia. 
 Dietary information  has been collected from each participant. 
 The main goal of this project is to examine the relationship between type 1 diabetes (T1D) associated autoantibody seropositivity (seroconverted), which is an indicator of T1D onset, and the infants' gut microbiome.
  In this project, seven out of 157 infants are seroconverted.

The dataset   contains a total of 55 microbial genera and 762 samples from 157 infants. 
A large collection of potential associations between relative abundances of microbial taxa and  covariates has  been previously discussed in \cite{tommi}. 
Among these associations, the most significant ones  link nationality and age  to 44 microbial genera. 
Due  to  moderate  sample size,  only limited evidence of  variations  of the  microbiome profile associated  with seroconversion has been reported.    

We present  analyses  based  on the proposed Bayesian model. The  set of covariates is composed by nationality, age, seroconversion and the interaction between age and nationality. We  want  to verify    consistency  of  our posterior  inference with the results discussed in the literature.  Additionally, we want to   quantify  the  uncertainty of the estimated relationship between seroconversion and microbial  compositions  in human gut.  

\subsection{Estimating the effects of age}
\begin{figure}[htbp]
\centering
\includegraphics[scale=0.55]{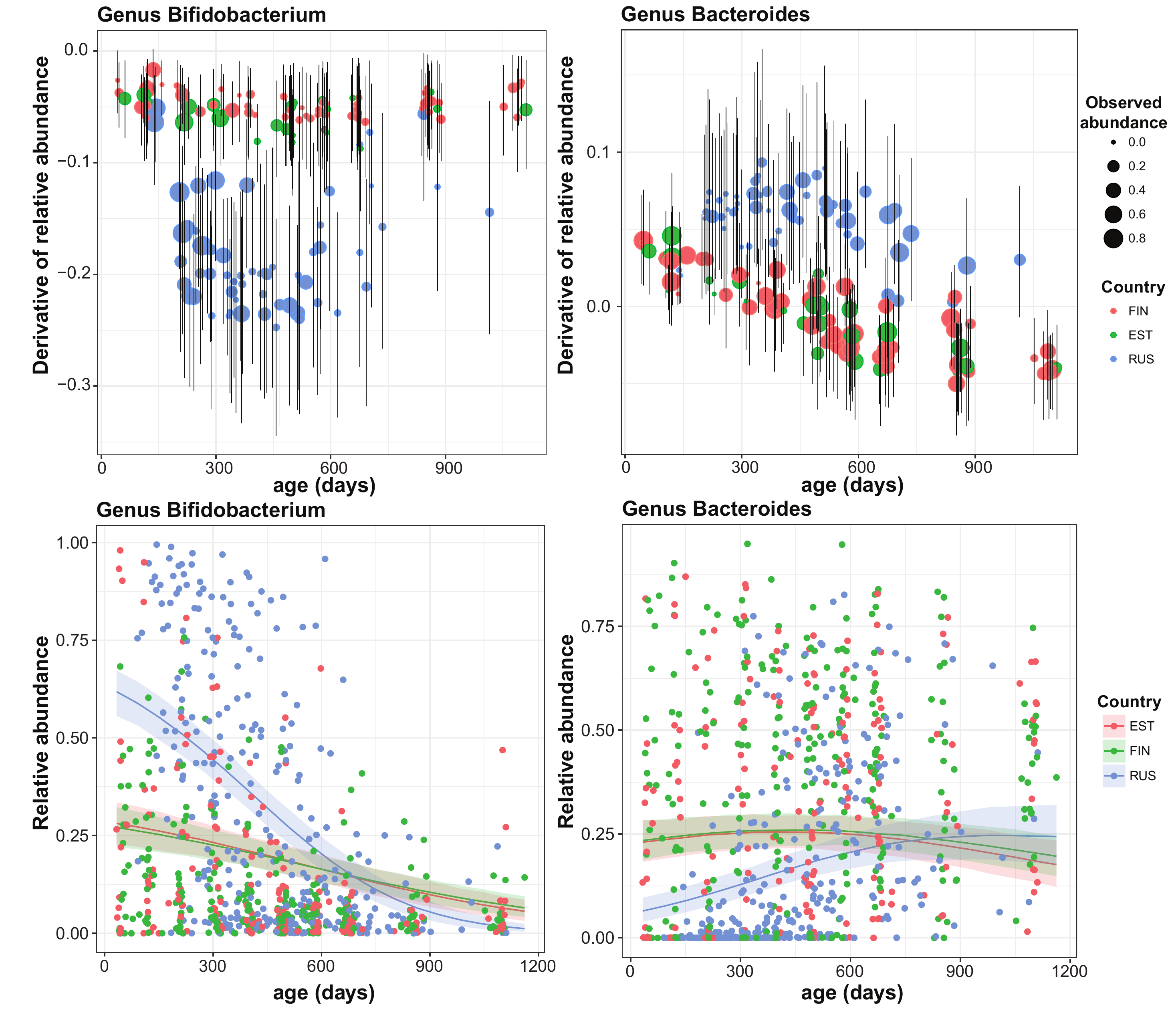}
\caption[
Estimated relationship between age and species abundances in DIABIMMUNE dataset]{(\textbf{Top}) Estimated $\partial P^j(\{Z_i\})/\partial w_{3,j}$ for two  genera. Each point represents a sample. Colors indicate nationalities and the sizes of points are proportional to the observed abundances.
 The error bars indicate  95\% credible intervals. We only plot 150 randomly selected samples.  (\textbf{Bottom}) Estimated population trends of Bifidobacterium and Bacteroides for Estonian, Finnish and Russian infants. The infants are assumed to be nonseroconverted. Curves represent the estimated population trends and the shaded areas illustrate  pointwise 95\% credible bands. Points indicate the observed abundance of Bifidobacterium or Bacteroides in all samples. We use colors to indicate nationalities.}
\label{bug.time}
\end{figure}

We  estimated the effects of age on microbial compositions using the  visualization approaches in Section \ref{comp.2}. 
In the top panels of Figure \ref{bug.time}, we illustrate the estimated derivatives of microbial abundances with respect to age, $\partial P^j(\{Z_i\})/\partial w_{3,j}$, for two  genera, Bifidobacterium and Bacteroides. We only plot the $\partial P^j(\{Z_i\})/\partial w_{3,j}$'s  for 150 randomly selected samples for visual clarity. We show the 95\% credible intervals for derivatives with  bars, and the sizes of points  are  proportional to the observed abundances. 

In the bottom panels of Figure \ref{bug.time}, we plot the estimated population trends of the  same genera with respect to age. 
We consider the population trends for Estonian, Finnish and Russian infants and assume that the infants are not seroconverted. 
Posterior credible bands for the population trends are visualized by shaded areas. 
The observed abundances of Bifidobacterium and Bacteroides in all samples are illustrated by scatter plots  together with the estimated population trends. 

The estimated derivatives with respect to age for Bifidobacterium are significantly smaller than zero in most of the samples, indicating that the abundances of Bifidobacterium in infants' gut microbiome tend to decrease with age. This is  to  some  extent   expected,  since bacteria from this genus are associated with breastfeeding \citep{fanaro2003intestinal}. The results on derivatives are consistent with the estimated population trends. In all three populations (Estonian, Finnish and Russian), Bifidobacterium is estimated to have a decreasing population trend with respect to age. The trends for Finnish and Estonian infants are similar, while for Russian infants the decrease is faster  for infants that are less than 600 days old. 

The association between genus Bacteroides and age is less pronounced. 
The derivatives of Bacteroides tend to be  positive in samples taken before 300 days.  
When the infants get older the derivatives become slightly negative in Estonian and Finnish infants but remain positive in the Russian  group. 
The population trends in this case are also consistent with the estimated derivatives. 
For nonseroconverted Estonian and Finnish infants, the estimated population  abundances of Bifidobactrium increase  with age when the infants are less than 450 days old and start to decrease slowly afterward.  In Russian infants, the initial increasing trend is more pronounced with a narrower credible band than the other two populations until 900 days. After 900 days, the population average abundance reaches a plateau and the credible band widens. 

\subsection{Estimating effects of nationalities and seroconversion}
\label{nation.effect}
\begin{figure}[htbp]
\centering
\includegraphics[scale=0.6]{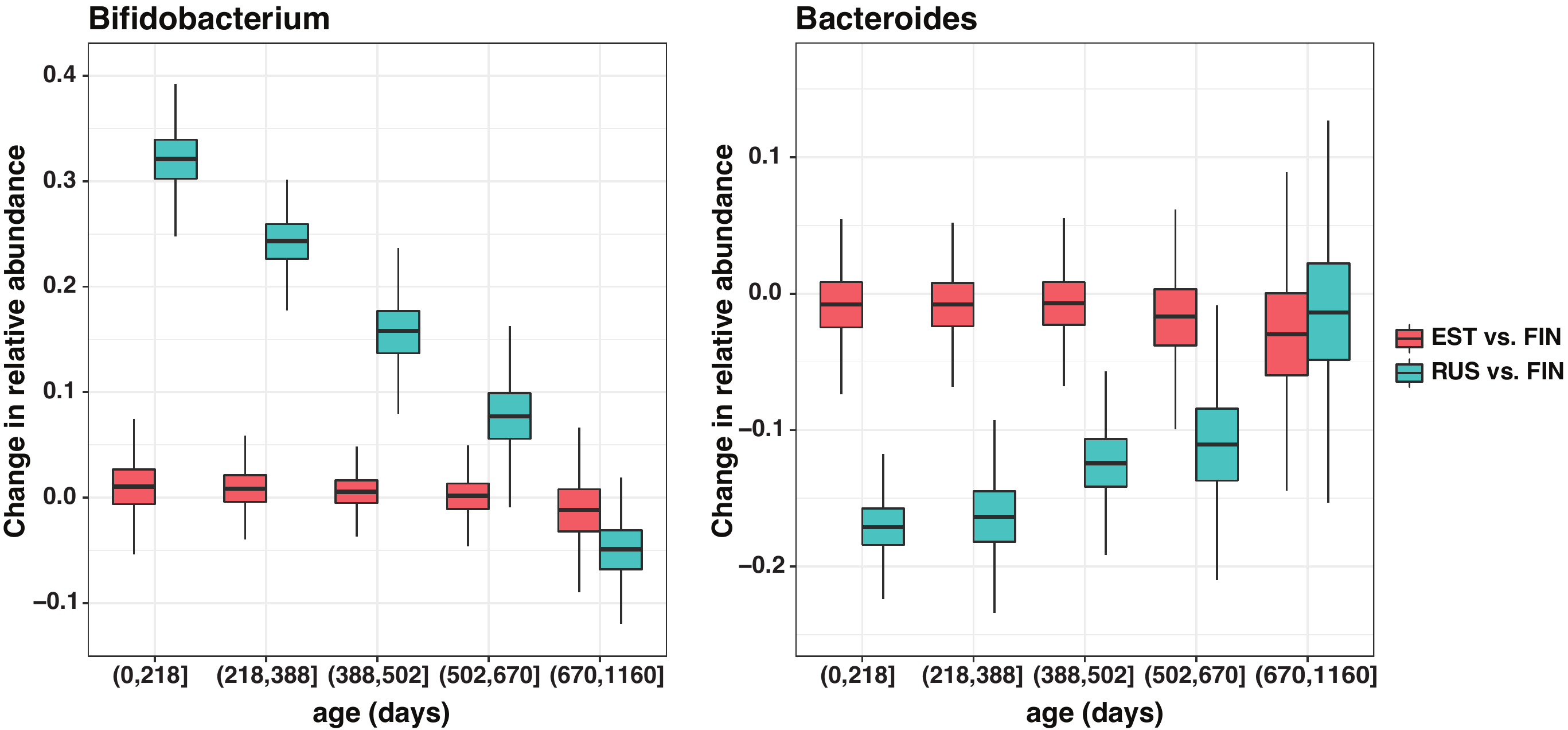}
\caption[Estimated relationship between country and species abundances in DIABIMMUNE dataset]{
Posterior distributions of  the average difference $\Delta P^j(\{Z_i\})/\Delta w_{1,j}$  (red) and $\Delta P^j(\{Z_i\})/\Delta w_{2,j}$ (green) in five consecutive age groups.  We plot the results for Bifidobacterium (\textbf{Left}) and Bacteroides (\textbf{Right}).}
\label{bug.c.s}
\end{figure}

We make inference about the associations between the gut microbial compositions and nationalities  using the differences $\Delta P^j(\{Z_i\})/\Delta w_{l,j}$ defined in (\ref{discrete.effect}). 
 For each sample, we estimate $\Delta P^j(\{Z_i\})/\Delta w_{1,j}$, which is the difference associated to the change of nationality from Finland (FIN) to Estonia (EST), as well as $\Delta P^j(\{Z_i\})/\Delta w_{2,j}$, the difference associated to the change from Finland (FIN) to Russia (RUS). We consider the averages of $\Delta P^j(\{Z_i\})/\Delta w_{1,j}$ and $\Delta P^j(\{Z_i\})/\Delta w_{2,j}$ in each of five consecutive age groups. The posterior distributions of these 
 population averages (Figure \ref{bug.c.s}) illustrate  the effect of nationality.
In both panels of Figure \ref{bug.c.s}, the X-axis identifies age groups and the Y-axis indicates the value of $\Delta P^j(\{Z_i\})/\Delta w_{1,j}$ and $\Delta P^j(\{Z_i\})/\Delta w_{2,j}$. Each box-plot approximates, using posterior simulations, the posterior distribution of the average $\Delta P^j(\{Z_i\})/\Delta w_{l,j},l=1,2$.
These averages  are defined by integration within a specific age  group. 

There is an  increase of Bifidobacterium abundance when  we compare  FIN to RUS nationalities. This increase diminishes  with age. 
In  the last age group (670-1160 days), the posterior distribution of  $\Delta P^j(\{Z_i\})/\Delta w_{2,j}$ indicates  that the abundances of Bifidobacterium in samples collected from infants older than 670 days remain comparable across nationalities. In the second comparison of nationalities, FIN to EST,  only minor  changes  of Bifidobacterium 
abundance levels  are observed. The abundances of Bacteroides are smaller in RUS than in FIN nationalities. This difference again diminishes with age. The difference of Bacteroides abundances between EST and FIN are also minor.

We also explored associations between microbial compositions and seroconversion status. In this case  we again examine the posterior distributions of average $\Delta P^j(\{Z_i\})/\Delta w_{4,j}$ in five consecutive age groups. 
We do not find evidence in our analyses of any  genus  associated  to seroconversion, due to high uncertainty of the estimated average $\Delta P^j(\{Z_i\})/\Delta w_{4,j}$ in all age groups.

\subsection{Similarities of microbial genera}

\begin{figure}[htbp]
\centering
\includegraphics[scale=0.45]{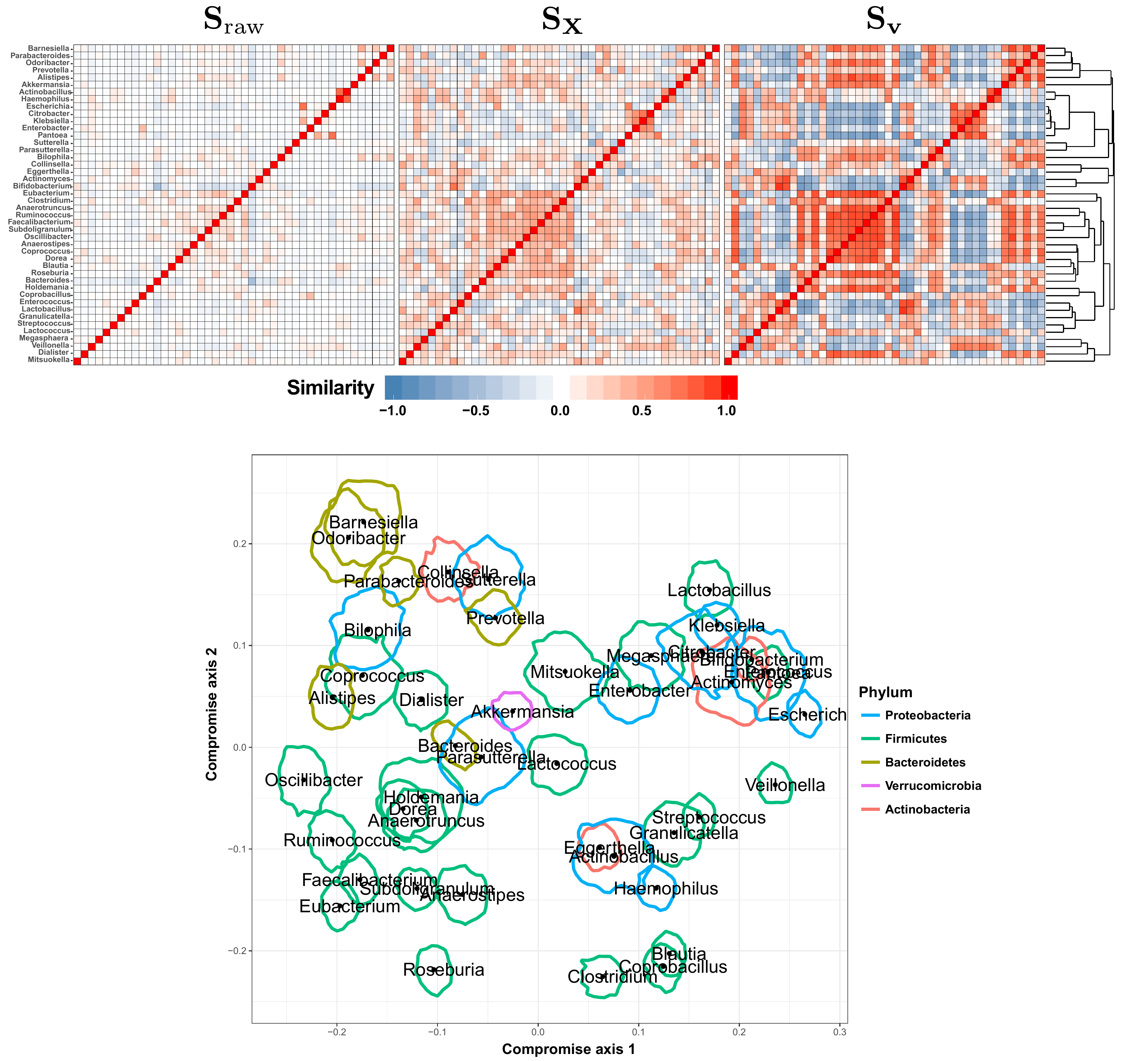}
\caption{{Estimated similarities of genera. (\textbf{Top}) Estimates of $\bS_{\bX}$, $\bS_{\bv}$ and $\bS_{\text{raw}}$. Each row or column in the heat-maps correspond to a specific genus. The color of each entry is determined by the estimated pair-wise similarity. The rows and columns in  heat-map are reordered so that adjacent rows or columns correspond to genera that are close phylogenetically. 
The phylogenetic tree for these genera are plotted at the right side of the figure. (\textbf{Bottom}) 
Ordination of genera based on $\bS_{\bX}$. The contour lines indicate  uncertainty regions  in the ordination configuration. The contour line of a genus is colored accordingly to the phylum of the genus.}}
\label{bug.bug}
\end{figure}

In this subsection, we focus on  similarities between microbial genera.
 We first consider the simple approach where the similarity of two genera is measured by the correlation between their observed relative abundances across all samples. 
 The result of this approach is a correlation matrix, denoted by $\bS_{\text{raw}}=(S_{\text{raw}}(i,i');i,i'\leq I)$, where $S_{\text{raw}}(i,i') = \text{cor}[(n_{i,j}/n_j;j\leq J),(n_{i',j}/n_j;j\leq J)]$. We then consider two approaches which utilize the proposed model.
  The first approach uses the cosine of the angle between $\bv_i$ and $\bv_{i'}$ to quantify the similarity of genera $i$ and $i'$, 
  whereas the second approach uses the cosine of the angle between $\bX_i$ and $\bX_{i'}$. The results of these two approaches are normalized Gram matrices, denoted as $\bS_{\bv}$ and $\bS_{\bX}$ respectively. In the top panels of Figure \ref{bug.bug}, we illustrate the estimates of $\bS_{\text{raw}}$, $\bS_{\bv}$ and $\bS_{\bX}$ by heat-maps. Each row or column of the heat-map represents a specific genus and the color of each tile represents the estimated similarity of two genera.

We then focus on examining the concordance of $\bS_{\text{raw}}$, $\bS_{\bv}$ and $\bS_{\bX}$ to the phylogenetic relations of the observed genera. To this end, we compare the phylogenetic tree of the observed genera published in \cite{segata2013phylophlan} to the heat-maps. If an  estimated correlation matrix indicates clusters of genera that share similarities with the phylogenetic tree, then we conclude that the estimate is consistent with phylogenetic relations.

From the figures we can find that $\bS_{\text{raw}}$ indicates little  between-genera similarity and  does not recover phylogenetic relations of the observed genera. On the other hand, both $\bS_{\bX}$ and $\bS_{\bv}$ indicate clusters of genera that are consistent with the phylogenetic tree. 
For instance, the cluster in the middle of the heat-maps of $\bS_{\bX}$ and $\bS_{\bv}$ corresponds to 13 genera from phylum Firmicutes (Clostridium, Ruminococcus, etc). 
These results suggest that both $\bS_{\bX}$ and $\bS_{\bv}$ capture the phylogenetic relations of the observed genera. The ordination plot of genera based on $\bS_{\bX}$ in the bottom panel of Figure \ref{bug.bug} further confirms this conclusion. We generate the ordination plot using the method in \cite{boyuren}, which represents each genus by a region instead of a single point.  In the ordination plot we find that genera from the same cluster in $\bS_{\bX}$ or $\bS_{\bv}$ are close to each other.

We also verify quantitatively the consistency of $\bS_{\bX}$ and $\bS_{\bv}$ to the phylogenetic relations. 
We first calculate the pair-wise phylogenetic distance matrix of the observed genera using unweighted-Unifrac dissimilarity \citep{lozupone2011unifrac}. We then convert this distance matrix into a normalized Gram matrix $\bS_{\text{unifrac}}$ by Torgerson Classical Scaling \citep{borg2005modern} and compare $\bS_{\text{unifrac}}$ to $\bS_{\text{raw}}$, $\bS_{\bX}$ and $\bS_{\bv}$. The estimated $\bS_{\bX}$ and $\bS_{\bv}$ are both similar to $\bS_{\text{unifrac}}$ with RV-coefficients 0.66 and 0.76 respectively, while the RV-coefficient between  $\bS_{\text{raw}}$ and $\bS_{\text{unifrac}}$ is 0.32.


\subsection{Goodness-of-fit of the model}
We conducted  goodness-of-fit analyses for our model based on the model evaluation approach proposed in Section \ref{dir.mimix.comp}, see for  example the results shown in Section S8 of the Supplementary Material. 
We use posterior predictive evaluations to examine whether the observed distributions of reads for the two species  discussed  in this section, Bacteroides and Bifidobacterium, are close to the corresponding posterior predictive distributions. We construct the leave-one-out 95\% posterior predictive intervals of the relative abundances of Bacteroides and Bifidobacterium in biological sample $j$ based on data with biological sample $j$ excluded. We then check if the leave-one-out posterior predictive intervals cover the observed abundances of Bacteroides and Bifidobacterium. In our case, the predictive intervals for 93.2\% of all biological samples cover the observed relative abundances of Bacteroides and 96.3\% of all biological samples for Bifidobacterium. The high proportions of coverage for both genera indicate that there is no systematic discrepancy between the observed data and the fitted model.

\section{Discussion}
\label{sec:6}
We proposed a  Bayesian mixed effects regression model to perform multivariate  analyses for microbiome data. This regression analysis estimates the effects of covariates on microbial composition while allowing for  correlations  of the residuals. We  illustrate that  the model parameters are identifiable. This result  is  consistent  with  our simulation study. The model allows us to infer the relationship between  covariates and microbial compositions  with   two visualization approaches.  
In  simulations  we  showed  that  both the individual-level and the population-level relationships between covariates and microbial compositions  can  be  accurately  estimated.
Moreover, our model is more robust against zero-inflation than a latent factor model based on logistic-normal distribution.
We finally applied  the model to a longitudinal microbiome dataset and  compared  our results with those  previously  reported in the literature.  

The current posterior computation is implemented with a  Gibbs-sampler. This can be inefficient  when the number of parameters is large. The computation  time  increases approximately linearly  with   the number  of  samples  and,  similarly, with  the number  of microbial species. For the longitudinal microbial dataset that we analyzed  the computation time of one chain with $10^5$ iterations is around 90 minutes. A possible  substantial improvement  in  computation  time   can probably  be  obtained  with  Hamiltonian Monte Carlo  or variational Bayes methods.

In the  future we would also like to investigate  appropriate variable selection techniques for the fixed effects.
 This is  particularly  helpful  in settings  with large collections of covariates. 
A more flexible model for the fixed effects is also desirable. Currently, the relationship between  abundances  and  covariates are depicted by  linear functions
of the samples  characteristics,   possibly  augmented  by  pre-specified  transformations  of  the covariates. 
Finally,   the current prior  specification ignores potential relationship  across  regression vectors  $\bv_i$ associated to similar microbial species.
 A systematic way to incorporate such information would involve the specification of  a prior distribution  on $\bv$  that  mirrors  the phylogeny of microbial species.

\section{Supplementary Materials}
We provide the proof of the proposition for model identifiability in the general setting. We also include additional supporting plots and tables for the simulation studies and data application.

\bibliographystyle{biom}
\bibliography{reference}

\begin{thebibliography}{}

\bibitem[\protect\citeauthoryear{Albert and Chib}{Albert and
  Chib}{1993}]{albert1993bayesian}
Albert, J.~H. and Chib, S. (1993).
\newblock Bayesian analysis of binary and polychotomous response data.
\newblock {\em Journal of the American statistical Association} {\bf 88,}
  669--679.

\bibitem[\protect\citeauthoryear{Anders and Huber}{Anders and
  Huber}{2010}]{deseq}
Anders, S. and Huber, W. (2010).
\newblock Differential expression analysis for sequence count data.
\newblock {\em Genome biology} {\bf 11,} R106.

\bibitem[\protect\citeauthoryear{Arbel, Mengersen, Rousseau,
  et~al\mbox{.}}{Arbel et~al.}{2016}]{arbel}
Arbel, J., Mengersen, K., Rousseau, J., et~al. (2016).
\newblock Bayesian nonparametric dependent model for partially replicated data:
  the influence of fuel spills on species diversity.
\newblock {\em The Annals of Applied Statistics} {\bf 10,} 1496--1516.

\bibitem[\protect\citeauthoryear{Bhattacharya and Dunson}{Bhattacharya and
  Dunson}{2011}]{dunsonfactor}
Bhattacharya, A. and Dunson, D.~B. (2011).
\newblock Sparse bayesian infinite factor models.
\newblock {\em Biometrika} {\bf 98,} 291--306.

\bibitem[\protect\citeauthoryear{Borg and Groenen}{Borg and
  Groenen}{2005}]{borg2005modern}
Borg, I. and Groenen, P.~J. (2005).
\newblock {\em Modern multidimensional scaling: Theory and applications}.
\newblock Springer Science \& Business Media.

\bibitem[\protect\citeauthoryear{Brooks and Gelman}{Brooks and
  Gelman}{1998}]{Rhat}
Brooks, S.~P. and Gelman, A. (1998).
\newblock General methods for monitoring convergence of iterative simulations.
\newblock {\em Journal of computational and graphical statistics} {\bf 7,}
  434--455.

\bibitem[\protect\citeauthoryear{Chen and Li}{Chen and
  Li}{2013}]{multinomial-dirichlet}
Chen, J. and Li, H. (2013).
\newblock Variable selection for sparse dirichlet-multinomial regression with
  an application to microbiome data analysis.
\newblock {\em The annals of applied statistics} {\bf 7,}.

\bibitem[\protect\citeauthoryear{Fanaro, Chierici, Guerrini, and Vigi}{Fanaro
  et~al.}{2003}]{fanaro2003intestinal}
Fanaro, S., Chierici, R., Guerrini, P., and Vigi, V. (2003).
\newblock Intestinal microflora in early infancy: composition and development.
\newblock {\em Acta paediatrica} {\bf 92,} 48--55.

\bibitem[\protect\citeauthoryear{Ferguson}{Ferguson}{1973}]{ferguson1973bayesian}
Ferguson, T.~S. (1973).
\newblock A bayesian analysis of some nonparametric problems.
\newblock {\em The annals of statistics} pages 209--230.

\bibitem[\protect\citeauthoryear{Gevers, Kugathasan, Denson, V{\'a}zquez-Baeza,
  Van~Treuren, Ren, Schwager, Knights, Song, Yassour, et~al\mbox{.}}{Gevers
  et~al.}{2014}]{gevers2014treatment}
Gevers, D., Kugathasan, S., Denson, L.~A., V{\'a}zquez-Baeza, Y., Van~Treuren,
  W., Ren, B., Schwager, E., Knights, D., Song, S.~J., Yassour, M., et~al.
  (2014).
\newblock The treatment-naive microbiome in new-onset crohn's disease.
\newblock {\em Cell host \& microbe} {\bf 15,} 382--392.

\bibitem[\protect\citeauthoryear{Grantham, Guan, Reich, Borer, and
  Gross}{Grantham et~al.}{2019}]{MIMIX}
Grantham, N.~S., Guan, Y., Reich, B.~J., Borer, E.~T., and Gross, K. (2019).
\newblock Mimix: A bayesian mixed-effects model for microbiome data from
  designed experiments.
\newblock {\em Journal of the American Statistical Association} pages 1--20.

\bibitem[\protect\citeauthoryear{Greenblum, Turnbaugh, and
  Borenstein}{Greenblum et~al.}{2012}]{greenblum2012metagenomic}
Greenblum, S., Turnbaugh, P.~J., and Borenstein, E. (2012).
\newblock Metagenomic systems biology of the human gut microbiome reveals
  topological shifts associated with obesity and inflammatory bowel disease.
\newblock {\em Proceedings of the National Academy of Sciences} {\bf 109,}
  594--599.

\bibitem[\protect\citeauthoryear{Griffin, Kolossiatis, and Steel}{Griffin
  et~al.}{2013}]{griffin2013comparing}
Griffin, J.~E., Kolossiatis, M., and Steel, M.~F. (2013).
\newblock Comparing distributions by using dependent normalized random-measure
  mixtures.
\newblock {\em Journal of the Royal Statistical Society: Series B (Statistical
  Methodology)} {\bf 75,} 499--529.

\bibitem[\protect\citeauthoryear{{Human Microbiome Project Consortium}}{{Human
  Microbiome Project Consortium}}{2012}]{HMP}
{Human Microbiome Project Consortium} (2012).
\newblock Structure, function and diversity of the healthy human microbiome.
\newblock {\em nature} {\bf 486,} 207.

\bibitem[\protect\citeauthoryear{Ishwaran and Zarepour}{Ishwaran and
  Zarepour}{2002}]{ishwaran2002exact}
Ishwaran, H. and Zarepour, M. (2002).
\newblock Exact and approximate sum representations for the dirichlet process.
\newblock {\em Canadian Journal of Statistics} {\bf 30,} 269--283.

\bibitem[\protect\citeauthoryear{James, Lijoi, and Pr{\"u}nster}{James
  et~al.}{2009}]{james-latent}
James, L.~F., Lijoi, A., and Pr{\"u}nster, I. (2009).
\newblock Posterior analysis for normalized random measures with independent
  increments.
\newblock {\em Scandinavian Journal of Statistics} {\bf 36,} 76--97.

\bibitem[\protect\citeauthoryear{Johnson, Ream, Towell, Williams, and
  Guerrero}{Johnson et~al.}{2013}]{johnson2013bayesian}
Johnson, D.~S., Ream, R.~R., Towell, R.~G., Williams, M.~T., and Guerrero, J.
  D.~L. (2013).
\newblock Bayesian clustering of animal abundance trends for inference and
  dimension reduction.
\newblock {\em Journal of agricultural, biological, and environmental
  statistics} {\bf 18,} 299--313.

\bibitem[\protect\citeauthoryear{Kostic, Gevers, Siljander, Vatanen,
  Hy{\"o}tyl{\"a}inen, H{\"a}m{\"a}l{\"a}inen, Peet, Tillmann, P{\"o}h{\"o},
  Mattila, et~al\mbox{.}}{Kostic et~al.}{2015}]{kostic2015dynamics}
Kostic, A.~D., Gevers, D., Siljander, H., Vatanen, T., Hy{\"o}tyl{\"a}inen, T.,
  H{\"a}m{\"a}l{\"a}inen, A.-M., Peet, A., Tillmann, V., P{\"o}h{\"o}, P.,
  Mattila, I., et~al. (2015).
\newblock The dynamics of the human infant gut microbiome in development and in
  progression toward type 1 diabetes.
\newblock {\em Cell host \& microbe} {\bf 17,} 260--273.

\bibitem[\protect\citeauthoryear{Ledoux and Talagrand}{Ledoux and
  Talagrand}{2013}]{slepian}
Ledoux, M. and Talagrand, M. (2013).
\newblock Probability in banach spaces: isoperimetry and processes.

\bibitem[\protect\citeauthoryear{Li}{Li}{2015}]{HZLi-review}
Li, H. (2015).
\newblock Microbiome, metagenomics, and high-dimensional compositional data
  analysis.
\newblock {\em Annual Review of Statistics and Its Application} {\bf 2,}
  73--94.

\bibitem[\protect\citeauthoryear{Lijoi, Nipoti, Pr{\"u}nster,
  et~al\mbox{.}}{Lijoi et~al.}{2014}]{lijoi2014bayesian}
Lijoi, A., Nipoti, B., Pr{\"u}nster, I., et~al. (2014).
\newblock Bayesian inference with dependent normalized completely random
  measures.
\newblock {\em Bernoulli} {\bf 20,} 1260--1291.

\bibitem[\protect\citeauthoryear{Lindley and Smith}{Lindley and
  Smith}{1972}]{lindley1972bayes}
Lindley, D.~V. and Smith, A.~F. (1972).
\newblock Bayes estimates for the linear model.
\newblock {\em Journal of the Royal Statistical Society. Series B
  (Methodological)} pages 1--41.

\bibitem[\protect\citeauthoryear{Lozupone, Lladser, Knights, Stombaugh, and
  Knight}{Lozupone et~al.}{2011}]{lozupone2011unifrac}
Lozupone, C., Lladser, M.~E., Knights, D., Stombaugh, J., and Knight, R.
  (2011).
\newblock Unifrac: an effective distance metric for microbial community
  comparison.
\newblock {\em The ISME journal} {\bf 5,} 169.

\bibitem[\protect\citeauthoryear{MacEachern}{MacEachern}{2000}]{DDP}
MacEachern, S.~N. (2000).
\newblock Dependent dirichlet processes.
\newblock {\em Unpublished manuscript, Department of Statistics, The Ohio State
  University} pages 1--40.

\bibitem[\protect\citeauthoryear{Morgan, Tickle, Sokol, Gevers, Devaney, Ward,
  Reyes, Shah, LeLeiko, Snapper, et~al\mbox{.}}{Morgan
  et~al.}{2012}]{xochi2012}
Morgan, X.~C., Tickle, T.~L., Sokol, H., Gevers, D., Devaney, K.~L., Ward,
  D.~V., Reyes, J.~A., Shah, S.~A., LeLeiko, N., Snapper, S.~B., et~al. (2012).
\newblock of the intestinal microbiome in inflammatory bowel disease and
  treatment.
\newblock {\em Genome biology} {\bf 13,} 1.

\bibitem[\protect\citeauthoryear{M{\"u}ller, Quintana, and Rosner}{M{\"u}ller
  et~al.}{2011}]{muller2011product}
M{\"u}ller, P., Quintana, F., and Rosner, G.~L. (2011).
\newblock A product partition model with regression on covariates.
\newblock {\em Journal of Computational and Graphical Statistics} {\bf 20,}
  260--278.

\bibitem[\protect\citeauthoryear{Paulson, Stine, Bravo, and Pop}{Paulson
  et~al.}{2013}]{metagenomeseq}
Paulson, J.~N., Stine, O.~C., Bravo, H.~C., and Pop, M. (2013).
\newblock Differential abundance analysis for microbial marker-gene surveys.
\newblock {\em Nature methods} {\bf 10,} 1200--1202.

\bibitem[\protect\citeauthoryear{Qin, Li, Raes, Arumugam, Burgdorf, Manichanh,
  Nielsen, Pons, Levenez, Yamada, et~al\mbox{.}}{Qin et~al.}{2010}]{MetaHIT}
Qin, J., Li, R., Raes, J., Arumugam, M., Burgdorf, K.~S., Manichanh, C.,
  Nielsen, T., Pons, N., Levenez, F., Yamada, T., et~al. (2010).
\newblock A human gut microbial gene catalogue established by metagenomic
  sequencing.
\newblock {\em nature} {\bf 464,} 59--65.

\bibitem[\protect\citeauthoryear{Quince, Lundin, Andreasson, Greco, Rafter,
  Talley, Agreus, Andersson, Engstrand, and D'amato}{Quince
  et~al.}{2013}]{quince2013impact}
Quince, C., Lundin, E.~E., Andreasson, A.~N., Greco, D., Rafter, J., Talley,
  N.~J., Agreus, L., Andersson, A.~F., Engstrand, L., and D'amato, M. (2013).
\newblock The impact of crohn's disease genes on healthy human gut microbiota:
  a pilot study.
\newblock {\em Gut} {\bf 62,} 952--954.

\bibitem[\protect\citeauthoryear{Ren, Bacallado, Favaro, Holmes, and
  Trippa}{Ren et~al.}{2016}]{boyuren}
Ren, B., Bacallado, S., Favaro, S., Holmes, S., and Trippa, L. (2016).
\newblock Bayesian nonparametric ordination for the analysis of microbial
  communities.
\newblock {\em arXiv preprint arXiv:1601.05156} .

\bibitem[\protect\citeauthoryear{Robert and Escoufier}{Robert and
  Escoufier}{1976}]{rvcoef}
Robert, P. and Escoufier, Y. (1976).
\newblock A unifying tool for linear multivariate statistical methods: the
  {RV}-coefficient.
\newblock {\em Applied statistics} pages 257--265.

\bibitem[\protect\citeauthoryear{Robinson, McCarthy, and Smyth}{Robinson
  et~al.}{2010}]{edger}
Robinson, M.~D., McCarthy, D.~J., and Smyth, G.~K. (2010).
\newblock edger: a bioconductor package for differential expression analysis of
  digital gene expression data.
\newblock {\em Bioinformatics} {\bf 26,} 139--140.

\bibitem[\protect\citeauthoryear{Rodriguez and Dunson}{Rodriguez and
  Dunson}{2011}]{rodriguez2011nonparametric}
Rodriguez, A. and Dunson, D.~B. (2011).
\newblock Nonparametric bayesian models through probit stick-breaking
  processes.
\newblock {\em Bayesian analysis (Online)} {\bf 6,}.

\bibitem[\protect\citeauthoryear{Segata, B{\"o}rnigen, Morgan, and
  Huttenhower}{Segata et~al.}{2013}]{segata2013phylophlan}
Segata, N., B{\"o}rnigen, D., Morgan, X.~C., and Huttenhower, C. (2013).
\newblock Phylophlan is a new method for improved phylogenetic and taxonomic
  placement of microbes.
\newblock {\em Nature communications} {\bf 4,} 2304.

\bibitem[\protect\citeauthoryear{Teh, Jordan, Beal, and Blei}{Teh
  et~al.}{2006}]{hdp}
Teh, Y.~W., Jordan, M.~I., Beal, M.~J., and Blei, D.~M. (2006).
\newblock Hierarchical dirichlet processes.
\newblock {\em Journal of the American Statistical Association} {\bf 101,}
  1566--1581.

\bibitem[\protect\citeauthoryear{Vatanen, Kostic, d’Hennezel, Siljander,
  Franzosa, Yassour, Kolde, Vlamakis, Arthur, H{\"a}m{\"a}l{\"a}inen,
  et~al\mbox{.}}{Vatanen et~al.}{2016}]{tommi}
Vatanen, T., Kostic, A.~D., d’Hennezel, E., Siljander, H., Franzosa, E.~A.,
  Yassour, M., Kolde, R., Vlamakis, H., Arthur, T.~D., H{\"a}m{\"a}l{\"a}inen,
  A.-M., et~al. (2016).
\newblock Variation in microbiome lps immunogenicity contributes to
  autoimmunity in humans.
\newblock {\em Cell} {\bf 165,} 842--853.

\bibitem[\protect\citeauthoryear{Vehtari, Gelman, and Gabry}{Vehtari
  et~al.}{2015}]{Vehtari1}
Vehtari, A., Gelman, A., and Gabry, J. (2015).
\newblock Pareto smoothed importance sampling.
\newblock {\em arXiv preprint arXiv:1507.02646} .

\bibitem[\protect\citeauthoryear{Vehtari, Gelman, and Gabry}{Vehtari
  et~al.}{2017}]{Vehtari2}
Vehtari, A., Gelman, A., and Gabry, J. (2017).
\newblock Practical bayesian model evaluation using leave-one-out
  cross-validation and waic.
\newblock {\em Statistics and Computing} {\bf 27,} 1413--1432.

\bibitem[\protect\citeauthoryear{Wadsworth, Argiento, Guindani, Galloway-Pena,
  Shelburne, and Vannucci}{Wadsworth et~al.}{2017}]{wadsworth2017integrative}
Wadsworth, W.~D., Argiento, R., Guindani, M., Galloway-Pena, J., Shelburne,
  S.~A., and Vannucci, M. (2017).
\newblock An integrative bayesian dirichlet-multinomial regression model for
  the analysis of taxonomic abundances in microbiome data.
\newblock {\em BMC bioinformatics} {\bf 18,} 94.

\bibitem[\protect\citeauthoryear{Xia, Chen, Fung, and Li}{Xia
  et~al.}{2013}]{logistic-normal}
Xia, F., Chen, J., Fung, W.~K., and Li, H. (2013).
\newblock A logistic normal multinomial regression model for microbiome
  compositional data analysis.
\newblock {\em Biometrics} {\bf 69,} 1053--1063.

\bibitem[\protect\citeauthoryear{Xu, Paterson, Turpin, and Xu}{Xu
  et~al.}{2015}]{xu2015}
Xu, L., Paterson, A.~D., Turpin, W., and Xu, W. (2015).
\newblock Assessment and selection of competing models for zero-inflated
  microbiome data.
\newblock {\em PloS one} {\bf 10,} e0129606.

\end{thebibliography}


\begin{thebibliography}{}

\bibitem[\protect\citeauthoryear{Brooks and Gelman}{Brooks and
  Gelman}{1998}]{Rhat}
Brooks, S.~P. and Gelman, A. (1998).
\newblock General methods for monitoring convergence of iterative simulations.
\newblock {\em Journal of computational and graphical statistics} {\bf 7,}
  434--455.

\bibitem[\protect\citeauthoryear{Geweke et~al\mbox{.}}{Geweke
  et~al.}{1991}]{geweke1991evaluating}
Geweke, J. et~al. (1991).
\newblock {\em Evaluating the accuracy of sampling-based approaches to the
  calculation of posterior moments}, volume 196.
\newblock Federal Reserve Bank of Minneapolis, Research Department Minneapolis,
  MN.

\bibitem[\protect\citeauthoryear{Hewitt and Savage}{Hewitt and
  Savage}{1955}]{hewitt1955symmetric}
Hewitt, E. and Savage, L.~J. (1955).
\newblock Symmetric measures on cartesian products.
\newblock {\em Transactions of the American Mathematical Society} {\bf 80,}
  470--501.

\bibitem[\protect\citeauthoryear{Ledoux and Talagrand}{Ledoux and
  Talagrand}{2013}]{slepian}
Ledoux, M. and Talagrand, M. (2013).
\newblock Probability in banach spaces: isoperimetry and processes.

\bibitem[\protect\citeauthoryear{Plummer, Best, Cowles, and Vines}{Plummer
  et~al.}{2006}]{CODA}
Plummer, M., Best, N., Cowles, K., and Vines, K. (2006).
\newblock Coda: convergence diagnosis and output analysis for mcmc.
\newblock {\em R news} {\bf 6,} 7--11.

\end{thebibliography}

\end{document}


\maketitle

\setcounter{figure}{0} 
\setcounter{table}{0} 
\section{Model identifiability in the general setting}
\label{iden.proof.sec}
\begin{prop}
\label{prop1}
Assume that $\bw_j$, $j=1,\ldots,J$, are independently  distributed with $\mathbb E(\bw_j\bw_j^\intercal)$ of full rank. 
 Consider two sets  of  parameters $(\bv,\bY,\bsigma) $ and $ (\bv',\bY',\bsigma')$  having  $\text{trace}(\bSigma)=\text{trace}(\bSigma')$,  where  $\bSigma  = \bY^\intercal\bY + \bI $ and symmetrically $\bSigma' = (\bY')^\intercal \bY' + \bI$. If $[\bv,\bY,(\sigma_i/\sigma_{i'};i\neq i')]$ and $[\bv',\bY',(\sigma'_i/\sigma'_{i'};i\neq i')]$ are  different,  then there exists  an  integer $n_0>0$, such that  when  $n_j\ge n_0$, $j=1,\ldots,J$, under the two sets of parameters, the joint distributions of  $[(n_{i,j};i\geq 1, j\leq J),\bw]$ are different.
\end{prop}

We note that the requirement $\text{trace}(\bSigma)=\text{trace}(\bSigma')$ is not restrictive in data analysis. This is because the model is invariant to scale transformation of $\bQ$. By scaling $\bQ$  we can make $\text{trace}(\bSigma)$ to be equal to any pre-specified value.

To prove the above proposition, we first show that for parameter values $(\bsigma,\bY,\bv)$ and $(\bsigma',\bY',\bv')$, the  
equality, under  the two sets  of  parameters, of  the joint  distribution of $[(P^j(\{Z_i\});i\ge 1,j\leq J),\bw]$
implies 
\begin{equation}
\sigma_i/\sigma_{i'} = \sigma_i'/\sigma_{i'}', i\neq i',\;\;\;\;
\bS = \bS',\;\;\;\;
\bv_i=\bv_i', i\ge 1.
\label{param.eq}
\end{equation}

\begin{enumerate}
\item Denote $\Ptilde^j(\{Z_i\})=\mathbb I(P^j(\{Z_i\})>0)$,
$$
p(\Ptilde^j(\{Z_i\})|\bsigma,\bv,\bw_j,\bY)f(\bw_j) = \Phi((1-2\Ptilde^j(\{Z_i\}))\bv_i^\intercal\bw_j/\sqrt{\bSigma_{j,j}})f(\bw_j),
$$
where $\Phi$ is the CDF of the standard normal distribution and $f(\bw_j)$ is the density of $\bw_j$. 
If  the  joint  distribution  of $[(P^j(\{Z_i\});i\ge 1,j\leq J),\bw]$  is  identical  under the  two sets  of parameters,
\begin{equation}
[(P^j(\{Z_i\});i\ge 1,j\leq J),\bw] \mid \bv,\bsigma,\bY \overset{d}{=} [(P^j(\{Z_i\});i\ge 1,j\leq J),\bw]\mid \bv',\bsigma',\bY',
\label{dist.eq}
\end{equation}
 it follows that $\bv_i^\intercal\bw_j\Sigma_{j,j}^{-1/2}=(\bv_i')^\intercal\bw_j(\Sigma'_{j,j})^{-1/2}$ almost surely. With the assumption that $\mathbb E(\bw_j\bw_j^\intercal)$ is of full rank, we get $\bv_i\Sigma^{-1/2}_{j,j}=\bv_i'(\Sigma_{j,j}')^{-1/2}$ for $i\geq 1$ and $j=1,\ldots,J$.
    
If  $\bv=\bzero$, it is straightforward to verify that (\ref{dist.eq}) implies $\bv=\bv'$. 
If $\bv_i\neq \bzero$, since we know that for any $j\neq j'$,
\begin{align*}
\bv_i\Sigma_{j,j}^{-1/2} = \bv_i'(\Sigma_{j,j}')^{-1/2},\;\;\;\;\;
\bv_i\Sigma_{j',j'}^{-1/2} = \bv_i'(\Sigma_{j',j'}')^{-1/2},
\end{align*}
we have $\Sigma_{j,j}/\Sigma_{j',j'}=\Sigma_{j,j}'/\Sigma_{j',j'}'.$ This equality, combined with the assumption that $\text{trace}(\bSigma)=\text{trace}(\bSigma')$, implies  that $\Sigma_{j,j}=\Sigma_{j,j}'$ for all $j\leq J$, and therefore $\bv_i=\bv_i'$ for all $i\ge 1$ if (\ref{dist.eq}) holds.

\item We then prove that  (\ref{dist.eq}) implies $\bS=\bS'$. 
We write
\begin{equation}
\begin{aligned}
&f(\bw_j)f(\bw_{j'})p(\Ptilde^j(\{Z_i\}),\Ptilde^{j'}(\{Z_i\})|\bw,\bsigma,\bv,\bY)= \\
&\;\;\;\;\;\;\;\;\;\;\;\;\;\;f(\bw_j)f(\bw_{j'})\int_{\Asc} \frac{1}{2\pi}(1-S_{j,j'}^2)^{-1/2}\exp\left(-\frac{1}{2}\bq^\intercal\bS_{j:j'}^{-1} \bq \right)d\bq,
\end{aligned}
\label{slepian.prob}
\end{equation}
where $\Asc=\Asc_j\times\Asc_{j'}$. $\Asc_j=(-\infty,\bv_i^\intercal\bw_j\Sigma_{j,j}^{-1/2}]$ if $\Ptilde^j(\{Z_i\})=0$ and $\Asc_j=[\bv_i^\intercal\bw_j\Sigma_{j,j}^{-1/2},\infty)$ if $\Ptilde^j(\{Z_i\})=1$.  Using Corollary 3.12 in \cite{slepian}, we know that  
the  probability  in (\ref{slepian.prob}) is  monotone with respect  to $S_{j,j'}$. Based on the previous  paragraphs, $\Asc=\Asc'$ for every $\bw$. Therefore (\ref{dist.eq}) implies that $S_{j,j'}=S_{j,j'}'$ for all $j,j'\leq J$. When $\bv\neq \bzero$, we further get $\bSigma = \bSigma'$.

\item  Finally, we prove that (\ref{dist.eq}) implies  $\sigma_i/\sigma_{i'} = \sigma'_i/\sigma'_{i'}$ for all $i\neq i'$. By construction, 
$$
\frac{P^j(\{Z_i\})}{P^j(\{Z_{i'}\})}=\frac{\sigma_i}{\sigma_{i'}}\frac{(Q_{i,j})_+^{2}}{(Q_{i',j})_+^{2}}.
$$
We use the convention that the ratio is zero whenever the denominator is zero.  By combining (\ref{dist.eq})  and the  previous paragraphs, the joint distribution of $((Q_{i,j})_+^{2}/(Q_{i',j})_+^{2},\bw_j)$ remains the same when the parameters values change from $(\bsigma,\bv,\bY)$ to $(\bsigma',\bv',\bY')$. This directly implies that $\sigma_i/\sigma_{i'}=\sigma_i'/\sigma_{i'}'$ for all $i\neq i'$ if (\ref{dist.eq}) holds.
\end{enumerate}

If (\ref{param.eq}) does not hold, then  the  joint  distribution of  $[(P^j(\{Z_i\});i\ge1, j\leq J),\bw]$  is  different  under the two sets  of parameters. By de Finetti's theorem \citep{hewitt1955symmetric}, the  joint distribution of the  observable variables $(n_{i,j};i\geq 1,j\leq J)$ and $\bw$ is uniquely determined by the mixing distribution, which in our case is the law of $(P^j(\{Z_i\});i\ge 1, j\leq J)|\bw,\bv,\bsigma,\bY$,  and  vice  versa. This  fact  completes  the proof.

\section{MCMC sampler for DirFactor}
\setcounter{figure}{0} 
\label{mixing.sec}
We focus on  identifiable components of our model. These are the normalized regression coefficients $v_{l,i}/\sqrt{\text{trace}(\bSigma)}$ and the correlation matrix $\bS$. We illustrate diagnostic summaries of  the MCMC for two  scenarios  (considered in Figure \ref{simple.mix} and Figure \ref{hard.mix}) defined in Section 4.3.
 In the first (simpler) scenario, the model is correctly specified 
 with no additional zero-inflation, $\var(\epsilon_{i,j})=1$
  and read depths generated  from a  Poisson distribution (see Section 4.3 for the description of this scenario).
  In the  second  scenario, the dataset is generated from MIMIX with a  moderate   addition of  zeros in the  microbial  compositions (Threshold$=10^{-3}$), $\var(\epsilon_{i,j})=10$ and read depths  are  generated  from a Negative binomial distribution with mean $10^5$ and variance $10^9$(see Section 4.3).
   
   For the correlation matrix, we illustrate the trace-plots of the first two eigenvalues. For the regression coefficients, we illustrate the trace-plots for $v_{1,1}/\sqrt{\text{trace}(\bSigma)}$ and $v_{1,2}/\sqrt{\text{trace}(\bSigma)}$. 
    We computed the upper confidence limit as recommended in \cite{CODA} of the $\hat R$ statistics \citep{Rhat} and performed Geweke's diagnostics \citep{geweke1991evaluating} for all  these parameters based on three MCMC chains to evaluate if the pre-specified number of Markov Chain transitions  is sufficient  for  approximate  posterior  inference.
        
    In the first simulation scenario, we consider 100,000 transitions for the Markov chains. The upper confidence limits of the $\hat R$ 
    statistics for the first two eigenvalues of $\bS$ and the scaled regression coefficients $\bv$ are all close to one (between 1.00 and 1.15), which suggests  that  approximately 60,000  MCMC iterations  might  be sufficient  for approximate  posterior  inference (Figure \ref{simple.mix}). This  is confirmed by Geweke's convergence diagnostic with truncation of the chains set at 60,000 iterations. The absolute Geweke's z-scores for all parameters and all chains are smaller than 0.5.
     
   In the second scenario, our model is misspecified and we consider longer Markov chains (200,000 transitions) to approximate posterior inference.
   The upper confidence limits of $\hat R$ for all parameters are close to one (between 1.00 and 1.05). The analyses  of the traceplots  suggest that approximately 100,000 iterations (Figure \ref{hard.mix}) might  be  sufficient  for  approximate  posterior inference. We verify this by calculating the absolute Geweke's z-scores for all four parameters across all chains when truncation of the chains is set at 100,000 iterations. All resulting absolute z-scores are smaller than 1.96. 
          
\begin{figure}[htbp]
\centering
\includegraphics[scale=0.45]{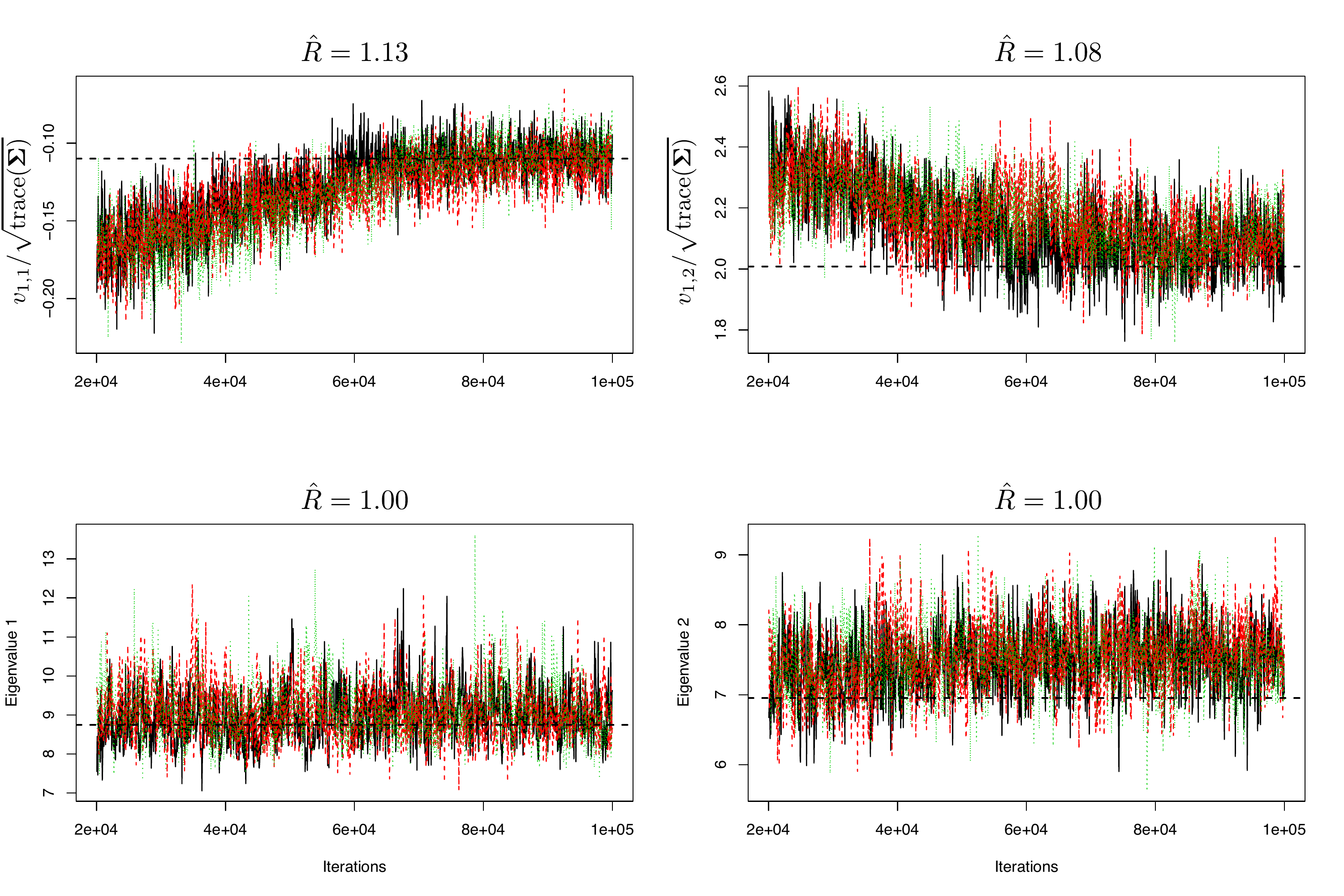}
\caption{ DirFactor  MCMC,  first simulation scenario. (\textbf{Top}) Trace-plots of two normalized regression coefficients $v_{1,1}/\sqrt{\text{trace}(\bSigma)}$ and $v_{1,2}/\sqrt{\text{trace}(\bSigma)}$. \textbf{(Bottom)} Trace-plots of the first two eigen-values of the correlation matrix $\bS$. The dash line in each figure indicates the actual value of the  parameter. The upper bounds of the $\hat R$
statistics based on the MCMC results from three chains are reported.}
\label{simple.mix}
\end{figure}

\begin{figure}[htbp]
\centering
\includegraphics[scale=0.45]{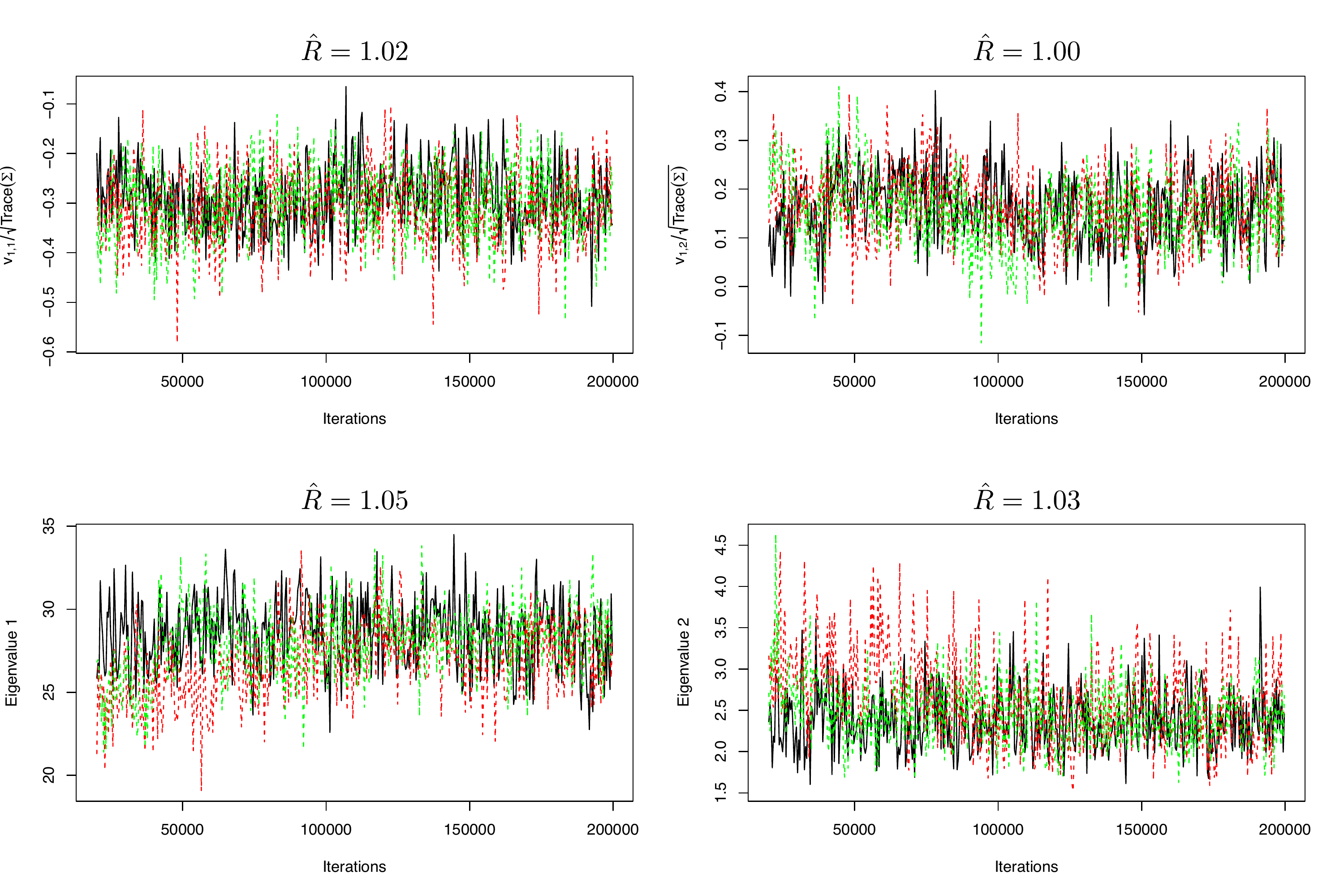}
\caption{DirFactor  MCMC,  second simulation scenario.(\textbf{Top})  Trace-plots of two normalized regression coefficients $v_{1,1}/\sqrt{\text{trace}(\bSigma)}$ and $v_{1,2}/\sqrt{\text{trace}(\bSigma)}$. \textbf{(Bottom)} Trace-plots of the first two eigen-values of the correlation matrix $\bS$. We do not visualize the actual value of the parameter since the data is not generated by our model. The upper bounds of the $\hat R$ statistics based on the MCMC results from three chains are reported.}
\label{hard.mix}
\end{figure}

\section{Permutation test}
\label{perm.test.sec}
We explore the use of the permutation procedure in Section 4.1 in a simulation study. 
The permutation test is computationally intensive, since we need  approximate  posterior  inference for  each permutation of the $w_{l,j}$ values.
In the  example  that we  describe, we used parallel computing. 
We consider the same simulation scenario  as in Section 4.1 and generate the data using our DirFactor model. 
As previously  mentioned we utilize  the posterior mean  $\hat \bv_l$ and  the Euclidean norm  $\|\hat\bv_l\|$  as  summary  statistics  for  testing.
We  generated  500 permutations and  derived a permutation-based  p-value   to  test the  null hypothesis $\bv_1=\bzero_I$. 
The  p-value  is smaller than 0.002, which indicates strong evidence of a  relation  between the  microbial  composition and the  covariate $l=1$.
 The test correctly detects the association between $w_{1,j}$ and the microbial composition. The null distribution of $\|\hat \bv_1\|$
 and the posterior mean $\|\hat\bv_1\|$  from the original data are shown in Figure \ref{perm.test}.

We also examine the behavior of the test under the null hypothesis, to evaluate if the Type-I error is controlled. We generate 50 datasets   from the same  scenario but we set $\bv_1=\bzero_I$. We then check if the distribution of the permutation p-values is uniform with a Q-Q plot (Figure \ref{perm.test}, right panel).
\setcounter{figure}{0} 
\begin{figure}[htbp]
\centering
\includegraphics[scale=0.5]{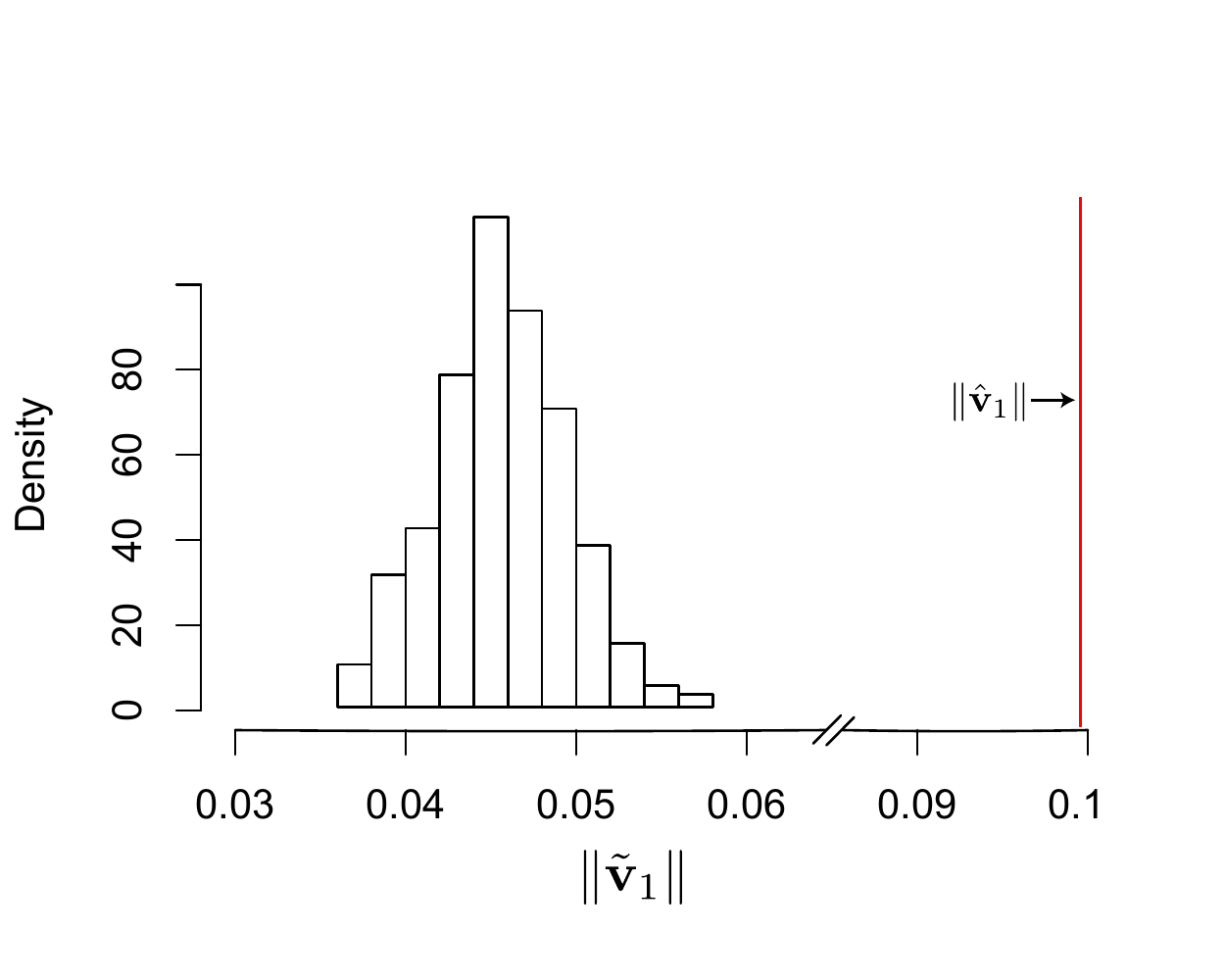}
\includegraphics[scale=0.5]{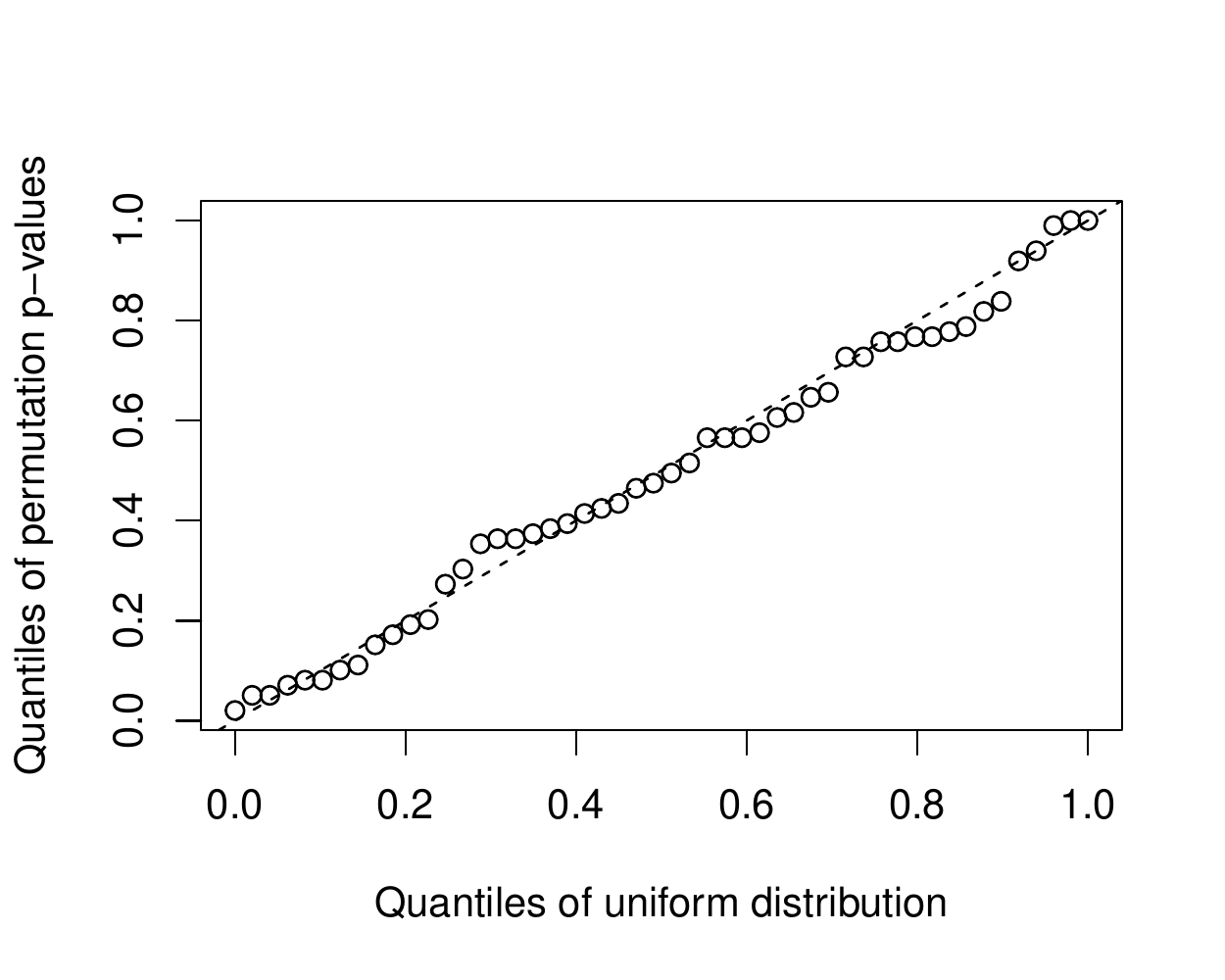}
\caption{(\textbf{Left}) Estimated null distribution of $\|\hat\bv_1\|$. The red line indicates the Euclidean norm of $\hat\bv_1$ from the original data. (\textbf{Right}) Q-Q plot compares the uniform distribution and the distribution of the permutation p-values for 50 independent datasets where $\bv_1=\bzero_I$.}
\label{perm.test}
\end{figure}

\section{Comparison of estimated derivatives to their actual values}
\label{deriv.comp.sec}
We illustrate additional  results  on the simulation study presented in  Section 4.2, with scatter plots that compare the estimated values of the partial derivatives $\partial P^j(\{Z_i\})/\partial w_{l,j}$ and their 95\% credible intervals to the  actual values. 
\setcounter{figure}{0}
\begin{figure}[htbp]
\centering
\includegraphics[scale=0.45]{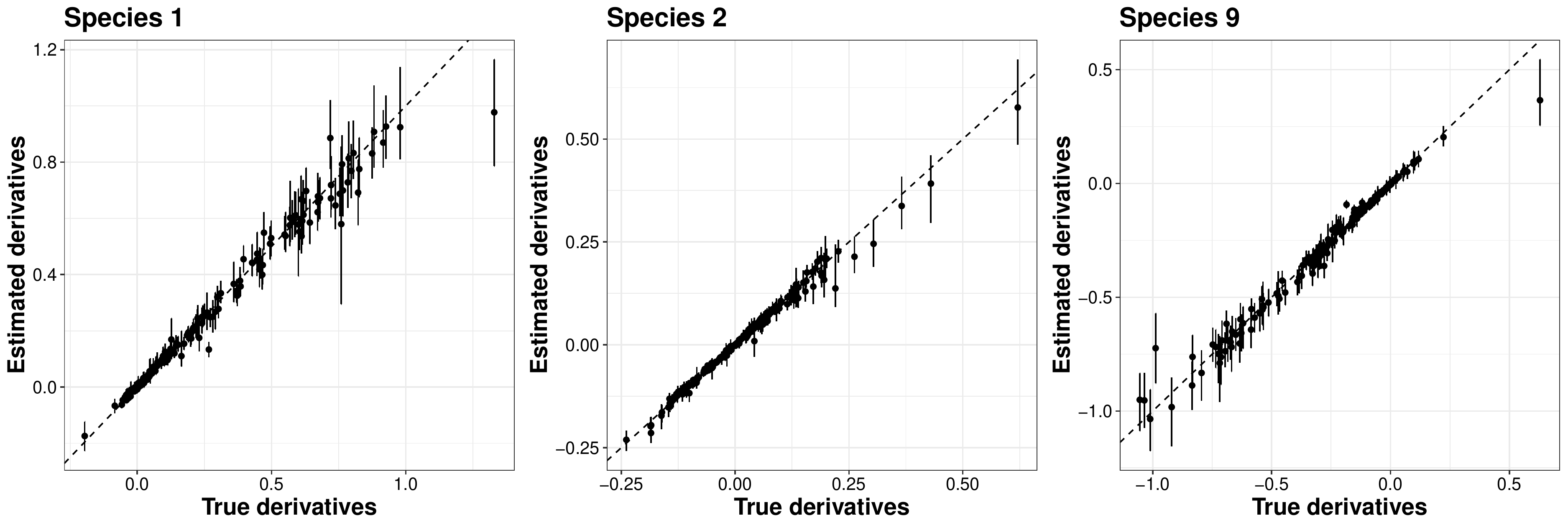}
\includegraphics[scale=0.45]{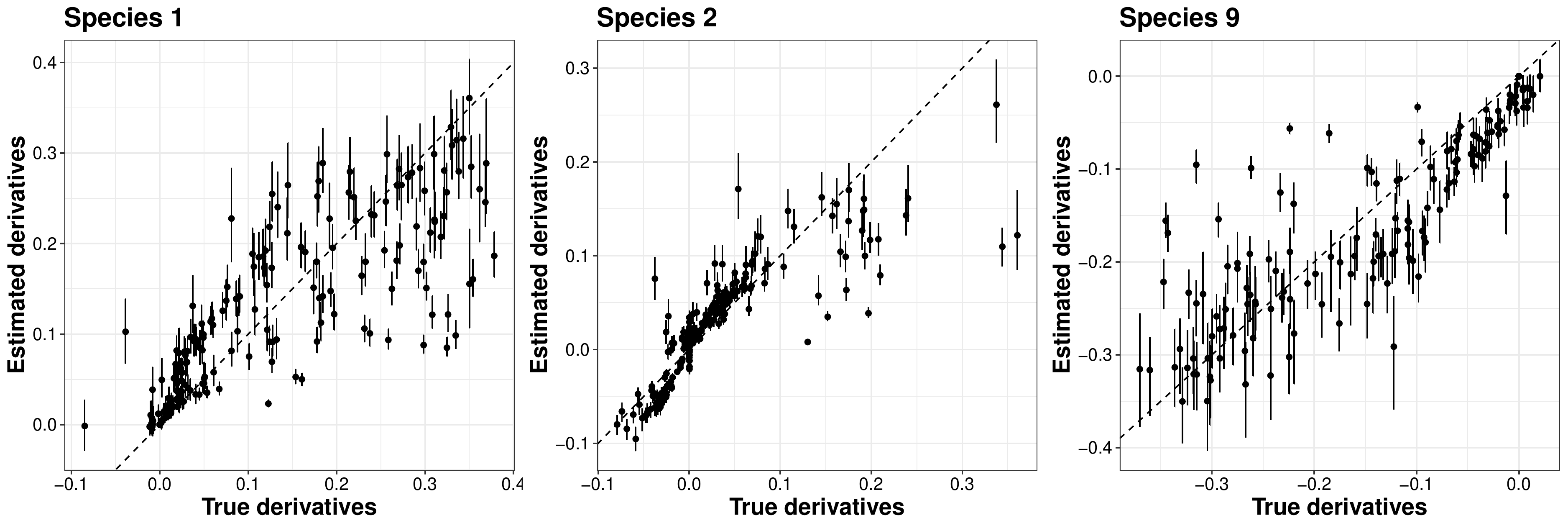}
\caption{Comparison between posterior estimates  of partial derivatives $\partial P^j(\{Z_i\})/\partial w_{l,j}$ 
and the corresponding actual values  when our model is correctly specified (\textbf{Top}) and misspecified (\textbf{Bottom}).  The  bar for each point indicates the 95\% credible interval of the partial derivative. Panels  include  a  45-degree  dashed line.}
\label{deriv.comp}
\end{figure}

\section{Derivatives and population trends in 50 simulation replicates}
\hspace{20mm}

\setcounter{figure}{0}
\begin{figure}[hb]
\centering
\includegraphics[scale=0.38]{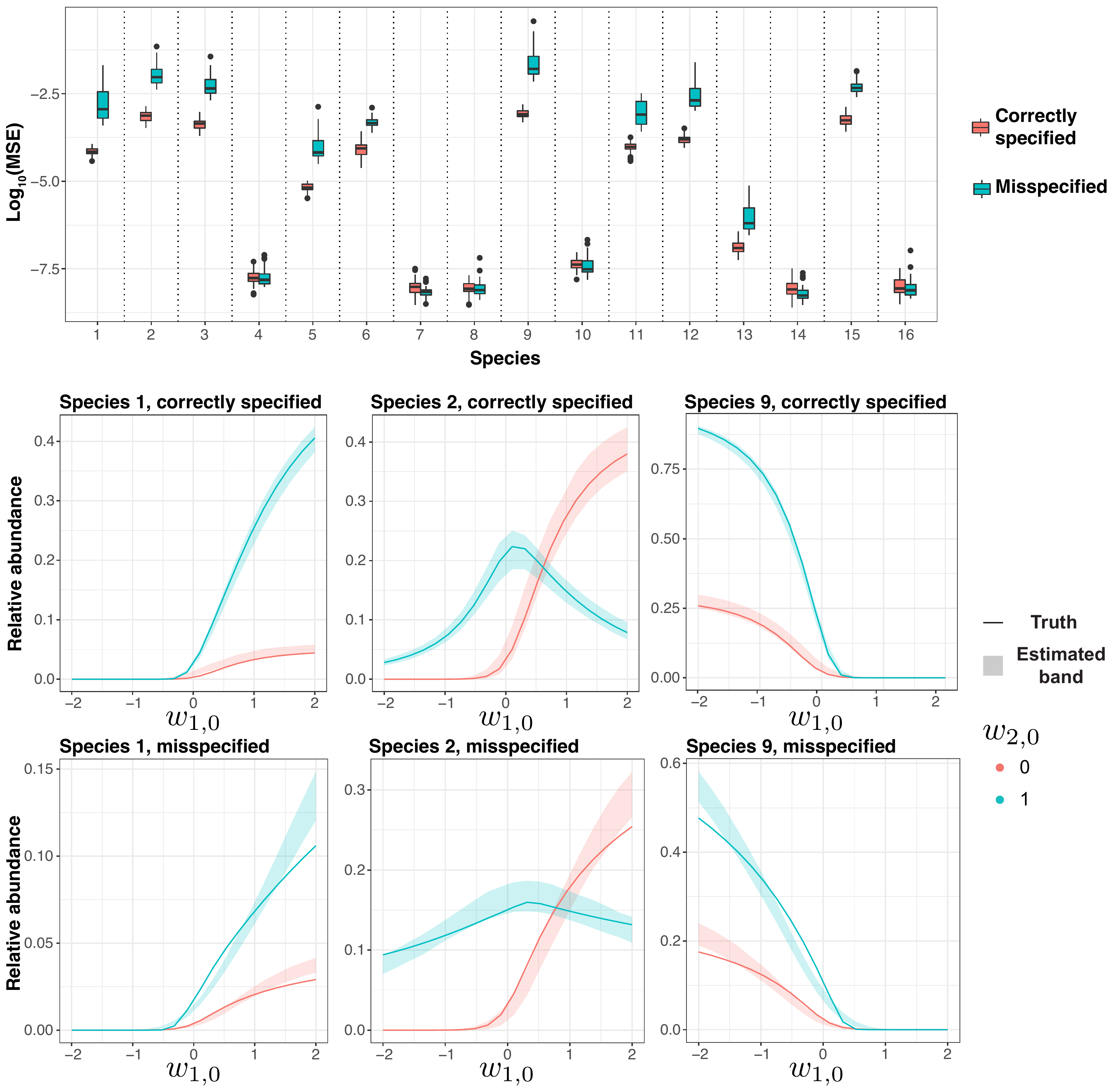}
\caption{Summary of the estimates of derivatives and population trends in 50 simulation replicates. (\textbf{Top}) We consider the distributions of MSEs of the estimated derivatives for the first 16 species, when the model is correctly specified and misspecified. (\textbf{Middle}) The  bands which include 
94\%  of  the  simulation-specific estimates of the population trends  along with the actual trends when the model is correctly specified. (\textbf{Bottom}) The same figures of population trends for the misspecified model.}
\label{dist_pw_mse}
\end{figure}

\section{Prediction accuracy and coverage}
\setcounter{figure}{0} 
\label{comp.res.read.depth.sec}
We summarize the results on prediction accuracy and coverage of DirFactor and MIMIX when $P^j(\{Z_i\})$'s are simulated as in Section 4.3 from either DirFactor or MIMIX and the read depths $n_j$ are generated from negative binomial distributions with moderate overdispersion (Table \ref{rmse.nb.1}, \ref{cov.nb.1}) and large overdispersion (Table \ref{rmse.nb.2}, \ref{cov.nb.2}).
We find that when the overdispersion of the read depths is large (mean $=10^5$ and variance $=4\times 10^{10}$), the prediction accuracy and coverage of both DirFactor and MIMIX decreases, while when the degree of overdispersion is moderate (mean $=10^5$ and variance $= 10^9$), the prediction accuracy and coverage remain  similar as in the simulations  where the read depths follow a Poisson distribution.

\begin{table}[htbp]
\centering
\begin{tabular}{c|ccc|ccc|ccc|ccc|}
      & \multicolumn{6}{c|}{Simulated from DirFactor} & \multicolumn{6}{c|}{Simulated from MIMIX} \bigstrut[b]\\
\cline{2-13}      & \multicolumn{3}{c|}{DirFactor} & \multicolumn{3}{c|}{MIMIX} & \multicolumn{3}{c|}{DirFactor} & \multicolumn{3}{c|}{MIMIX} \bigstrut\\
\hline
Threshold     & 0 & $10^{-3}$ & $10^{-2}$ & 0 & $10^{-3}$ & $10^{-2}$ & 0 & $10^{-3}$ & $10^{-2}$ & 0 & $10^{-3}$ & $10^{-2}$ \bigstrut[t]\\
\hline
$\var(\epsilon_{i,j})=1$ & 1.0&1.0&1.3&14.3&26.1&54.8&1.4&2.5&4.9&1.2&2.0&2.8\\
$\var(\epsilon_{i,j})=5$ &3.0&3.4&3.4&34.3&66.6&75.3&4.7&7.3&7.1&1.1&2.1&2.4\\
$\var(\epsilon_{i,j})=10$ &4.4&4.3&4.6&56.4&105.3&155.8&10.9&10.9&11.3&1.8&2.8&3.6\bigstrut[b]\\
\hline
\end{tabular}%
\caption{Average RMSE of estimated population mean abundances at 20 different values of $w_{1,0}$ equally spaced between -2 and 2 across simulation replicates for our model (DirFactor) and MIMIX. The threshold parameter indicates at which value we truncate the simulated $P^j(\{Z_i\})$'s to zero. We consider two scenarios where the dataset is generated from DirFactor and MIMIX respectively. The read depths are generated from a negative binomial distribution with mean $10^5$ and variance $10^9$. All RMSEs in the table are multiplied by $10^3$.}
\label{rmse.nb.1}
\end{table}

\begin{table}[htbp]
\centering
\begin{tabular}{c|ccc|ccc|ccc|ccc|}
      & \multicolumn{6}{c|}{Simulated from DirFactor} & \multicolumn{6}{c|}{Simulated from MIMIX} \bigstrut[b]\\
\cline{2-13}      & \multicolumn{3}{c|}{DirFactor} & \multicolumn{3}{c|}{MIMIX} & \multicolumn{3}{c|}{DirFactor} & \multicolumn{3}{c|}{MIMIX} \bigstrut\\
\hline
Threshold     & 0 & $10^{-3}$ & $10^{-2}$ & 0 & $10^{-3}$ & $10^{-2}$ & 0 & $10^{-3}$ & $10^{-2}$ & 0 & $10^{-3}$ & $10^{-2}$ \bigstrut[t]\\
\hline
$\var(\epsilon_{i,j})=1$&0.99&0.94&0.91&0.92&0.84&0.79&0.97&0.97&0.89&1.00&0.94&0.87\\
$\var(\epsilon_{i,j})=5$&0.99&0.93&0.88&0.94&0.84&0.80&0.96&0.93&0.89&0.99&0.93&0.84\\
$\var(\epsilon_{i,j})=10$&0.96&0.95&0.89&0.93&0.82&0.75&0.90&0.88&0.84&0.95&0.93&0.85\bigstrut[b]\\
\hline
\end{tabular}
\caption{Coverage of the posterior distribution of the  population trend (defined in Sec 3.2). We average across  species and across  $w_{1,0}$ values between -2 and  2  and  $w_{2,0}=0$ and $w_{2,0}=1$.  The threshold parameter indicates at which value we truncate the simulated $P^j(\{Z_i\})$'s to zero. The coverage is calculated using simulation replicates. The read depths are generated from a negative binomial distribution with mean $10^5$ and variance $10^9$. We consider two scenarios where the dataset is generated from DirFactor and MIMIX respectively.}
 \label{cov.nb.1}
\end{table}

\begin{table}[htbp]
\centering
\begin{tabular}{c|ccc|ccc|ccc|ccc|}
      & \multicolumn{6}{c|}{Simulated from DirFactor} & \multicolumn{6}{c|}{Simulated from MIMIX} \bigstrut[b]\\
\cline{2-13}      & \multicolumn{3}{c|}{DirFactor} & \multicolumn{3}{c|}{MIMIX} & \multicolumn{3}{c|}{DirFactor} & \multicolumn{3}{c|}{MIMIX} \bigstrut\\
\hline
Threshold     & 0 & $10^{-3}$ & $10^{-2}$ & 0 & $10^{-3}$ & $10^{-2}$ & 0 & $10^{-3}$ & $10^{-2}$ & 0 & $10^{-3}$ & $10^{-2}$ \bigstrut[t]\\
\hline
$\var(\epsilon_{i,j})=1$&2.2&3.0&4.5&24.4&35.9&53.0&2.2&4.3&6.1&2.2&2.7&2.9\\
$\var(\epsilon_{i,j})=5$&7.0&8.4&11.1&45.8&85.2&95.7&6.1&7.7&9.2&3.2&3.5&6.4\\
$\var(\epsilon_{i,j})=10$&8.4&10.5&12.6&75.5&126.0&185.6&15.7&18.7&21.7&4.6&6.3&7.2\bigstrut[b]\\
\hline
\end{tabular}%
\caption{Average RMSE of estimated population mean abundances at 20 different values of $w_{1,0}$ equally spaced between -2 and 2 across simulation replicates for our model (DirFactor) and MIMIX. The threshold parameter indicates at which value we truncate the simulated $P^j(\{Z_i\})$'s to zero. We consider two scenarios where the dataset is generated from DirFactor and MIMIX respectively. The read depths are generated from a negative binomial distribution with mean $10^5$ and variance $4\times 10^{10}$. All RMSEs in the table are multiplied by $10^3$.}
\label{rmse.nb.2}
\end{table}

\begin{table}[htbp]
\centering
\begin{tabular}{c|ccc|ccc|ccc|ccc|}
      & \multicolumn{6}{c|}{Simulated from DirFactor} & \multicolumn{6}{c|}{Simulated from MIMIX} \bigstrut[b]\\
\cline{2-13}      & \multicolumn{3}{c|}{DirFactor} & \multicolumn{3}{c|}{MIMIX} & \multicolumn{3}{c|}{DirFactor} & \multicolumn{3}{c|}{MIMIX} \bigstrut\\
\hline
Threshold     & 0 & $10^{-3}$ & $10^{-2}$ & 0 & $10^{-3}$ & $10^{-2}$ & 0 & $10^{-3}$ & $10^{-2}$ & 0 & $10^{-3}$ & $10^{-2}$ \bigstrut[t]\\
\hline
$\var(\epsilon_{i,j})=1$&0.98&0.81&0.72&0.96&0.61&0.49&0.99&0.86&0.76&0.96&0.77&0.70\\
$\var(\epsilon_{i,j})=5$&0.95&0.8&0.69&0.98&0.62&0.49&0.96&0.83&0.66&0.94&0.72&0.69\\
$\var(\epsilon_{i,j})=10$&0.9&0.77&0.65&0.99&0.64&0.5&0.90&0.8&0.65&0.94&0.72&0.64\bigstrut[b]\\
\hline
\end{tabular}
\caption{Coverage of the posterior distribution of the  population trend (defined in Sec 3.2). We average across  species and across  $w_{1,0}$ values between -2 and  2  and  $w_{2,0}=0$ and $w_{2,0}=1$.  The threshold parameter indicates at which value we truncate the simulated $P^j(\{Z_i\})$'s to zero. The coverage is calculated using simulation replicates. The read depths are generated from a negative binomial distribution with mean $10^5$ and variance $4\times 10^{10}$. We consider two scenarios where the dataset is generated from DirFactor and MIMIX respectively.}
 \label{cov.nb.2}
\end{table}

\section{Model comparison}
\setcounter{figure}{0} 
We use the same setting as in Section 4.3 without additional zero-inflation, Poisson distributed read-depths and $\var(\epsilon_{i,j})=1$ for the simulation of $Q_{i,j}$. After simulation of the $Q_{i,j}$ variables, we  use  them to generate two datasets, 
one consistent with  the distribution of our model ($P^j(\{Z_i\})\propto \sigma_i(Q_{i,j})_+^2$) and the other  consistent  with  MIMIX ($P^j(\{Z_i\})\propto \exp(Q_{i,j})$). 

\begin{figure}[htbp]
\centering
\includegraphics[scale=0.5]{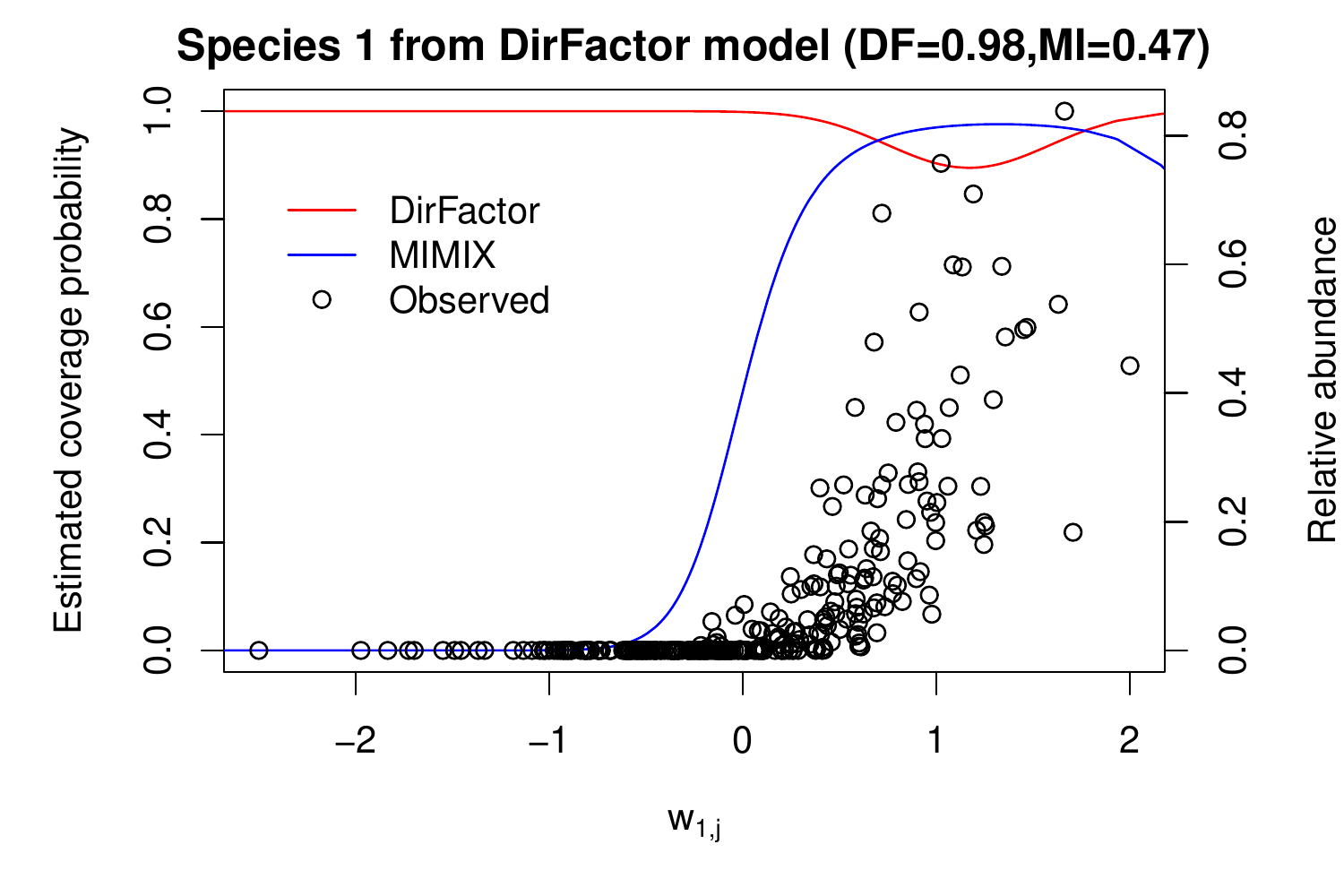}
\includegraphics[scale=0.5]{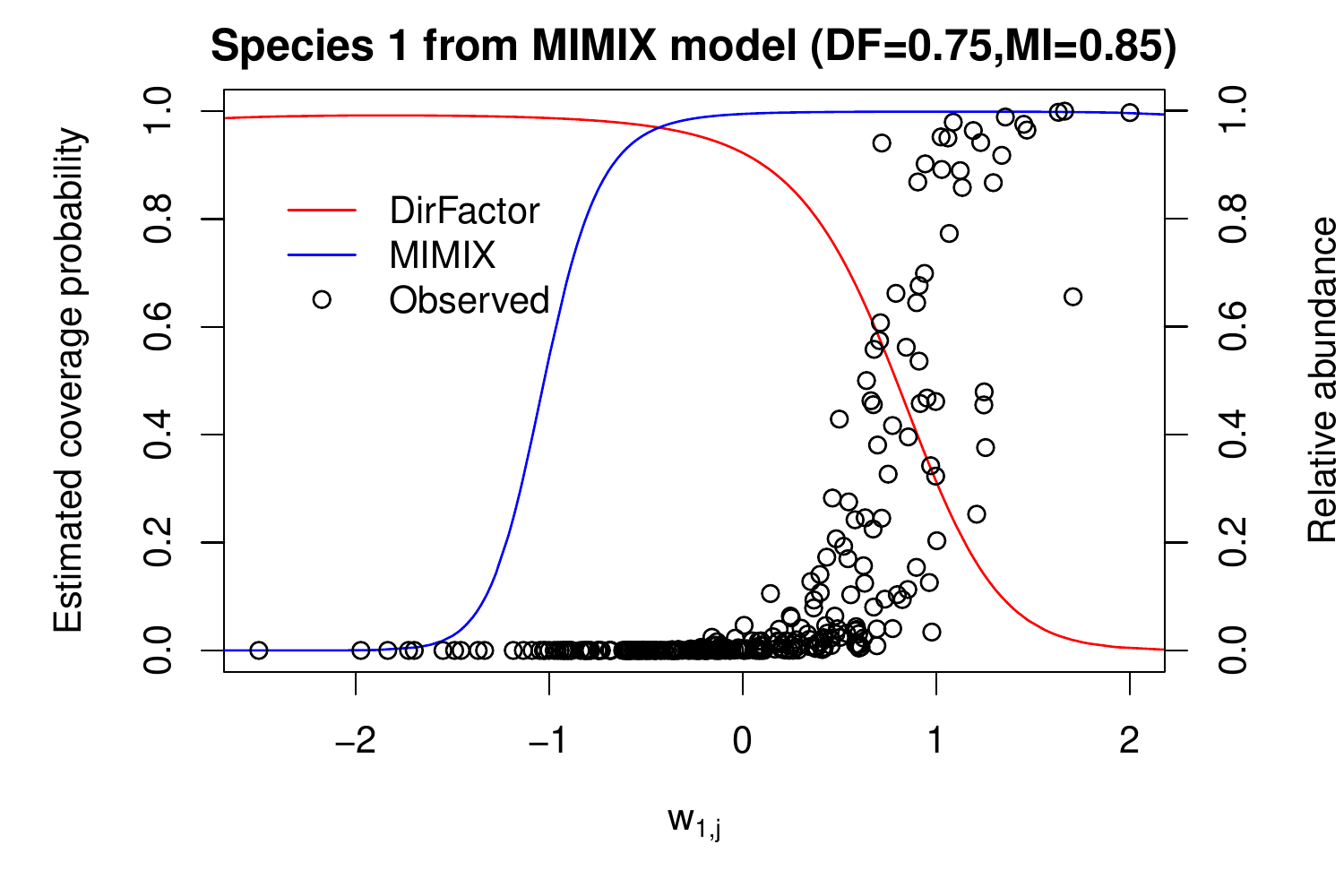}
\caption{Estimated coverage probabilities of posterior leave-one-out 95\% predictive intervals as a function of $w_{1,j}$ values. The curves indicate  estimated coverage probabilities and the points  illustrate the observed relative abundances of all samples. 
The curves  are obtained  by  fitting a binary regression model to  the covered or uncovered  status  of  each sample.
The color indicates which model is used to generated the posterior predictive intervals. The data are generated from DirFactor model (Left) and MIMIX model (Right). The two summaries  indicated as ``DF'' and ``MI'' are the proportions of samples that are covered by the predictive intervals generated by our Dirichlet model and the MIMIX model.}
\label{pp.sim.check}
\end{figure}

We  perform  posterior predictive evaluations of our model and MIMIX as outlined in Section 4.3. We use binary regression to illustrate the coverage probability of the 95\% predictive intervals as a function of $w_{1,j}$ for Species 1. An example is  visualized in Figure \ref{pp.sim.check}. The mean coverage probabilities, for  data  generated from MIMIX,  are 0.98 and 0.97 with  predictive  analyses  based  on MIMIX  and  our  model. Also, the mean coverage probabilities, for  data  generated from our model,  are 0.92 and 0.99  with  predictive  analyses  based  on MIMIX  and  our  model.

\section{Goodness-of-fit analysis for our model on the DIABIIMMUNE data}
\hspace{20mm}
\setcounter{figure}{0}
\begin{figure}[htbp]
\centering
\includegraphics[scale=0.5]{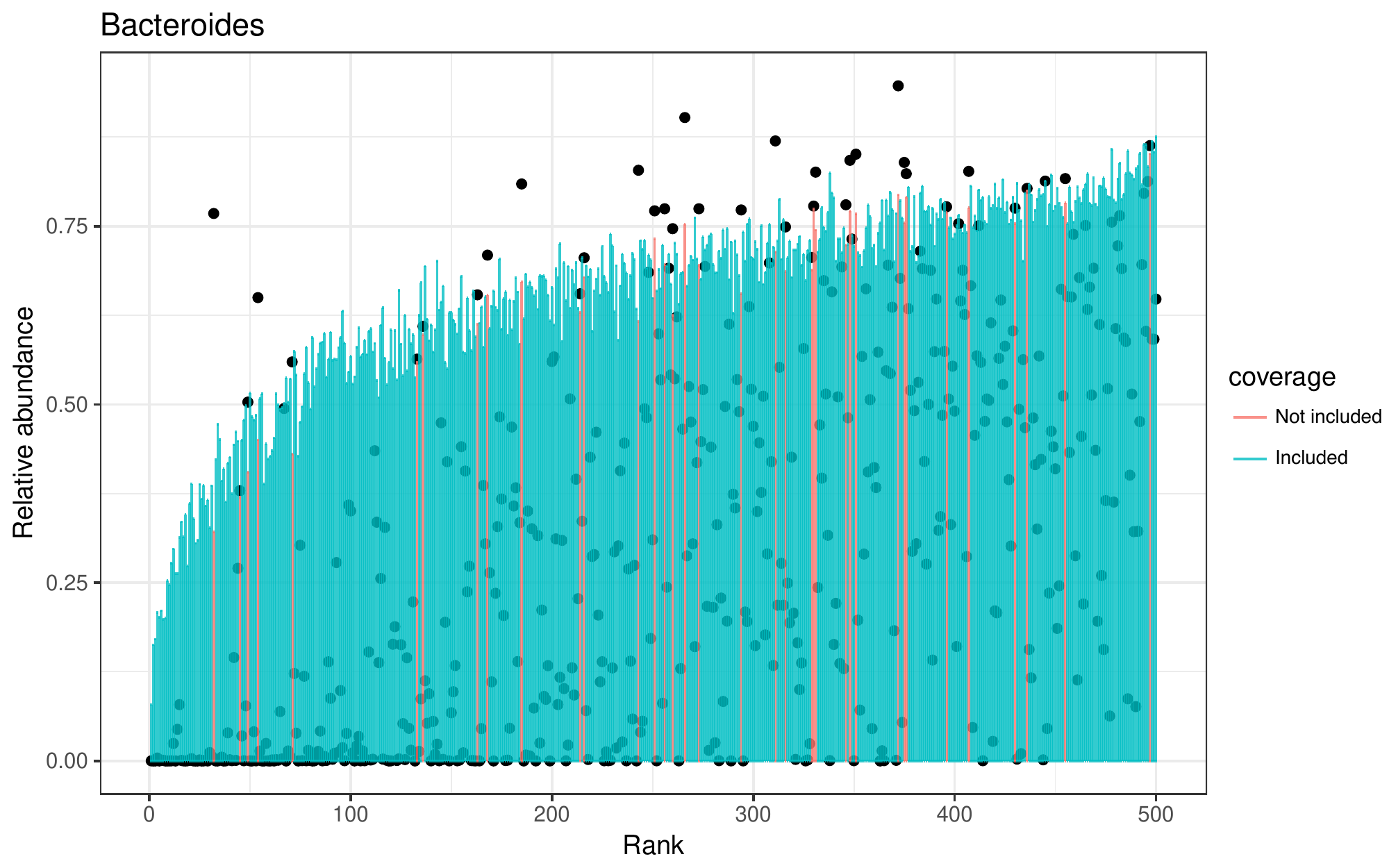}
\includegraphics[scale=0.5]{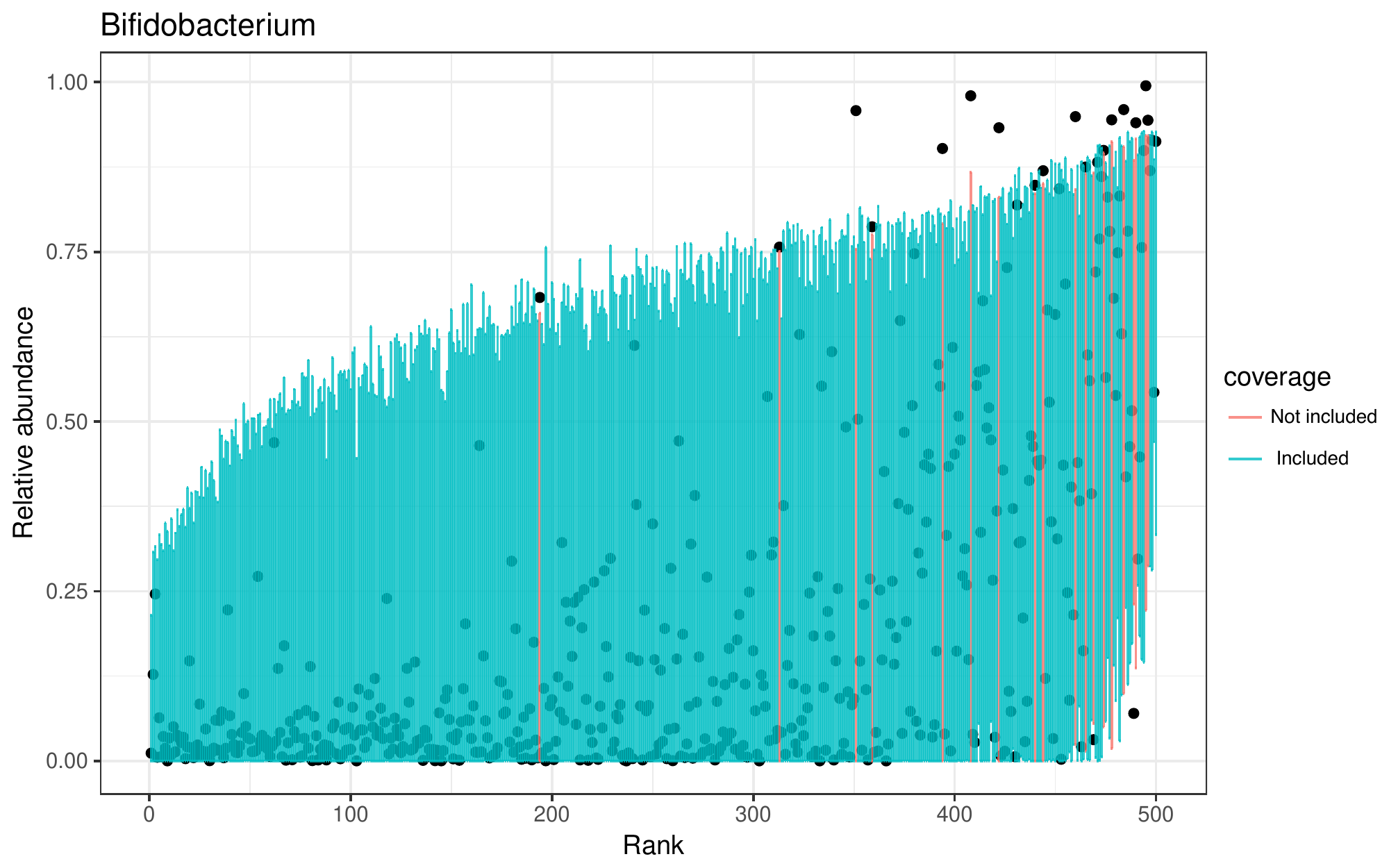}
\caption{Leave-one-out posterior predictive intervals for the abundances of Bacteroides and Bifidobactrium. We only illustrate 150 randomly sampled biological samples in this plot. Black dots indicate the observed relative abundance and lines indicate 95\% posterior predictive intervals. The biological samples are sorted by the posterior means of the associated predictive distributions of the relative abundances.}
\label{goodness.of.fit}
\end{figure}

\bibliographystyle{biom}
\bibliography{reference}